# First steps towards a theory of the Dense Plasma Focus:
## Part-I: Kinematic framework with built-in propagation delay and nonzero thickness of dense sheath for generalized electrode geometry

S K H Auluck

International Scientific Committee for Dense Magnetized Plasmas,
http://www.icdmp.pl/isc-dmp
Hery 23, P.O. Box 49, 00-908 Warsaw, Poland

## Abstract

This paper, Part I of a series, describes a kinematic framework for the theory of a Dense Plasma Focus which is very similar to the GV model in spirit but which differs in its scope in four respects. First, the GV model derives most of its results from the mathematical properties of the solution of a certain partial differential equation derived from assumptions that apparently represent conservation of momentum but are not a rigorous application of the relevant physics. The present model is based on the scaling properties of the standard equations of motion, which lead to mathematical results identical with the GV model. Second, the GV model is purely kinematic in nature. The present model is also kinematic like the GV model but it incorporates additional insights borrowed from other physical theories, models and experiments. Third, the GV model does not take into account the experimentally observed delay between the start of current and start of plasma propagation and the existence of a nonzero thickness of the dense plasma sheath. The present model incorporates both these features in its kinematic structure. Fourth, the unlike the GV model, the present model allows considerations of some modifications of standard Mather type geometry. In addition to the current waveform, the proposed model reproduces the height and radius of the pinch column, the ratio of pinch density to fill density, the general appearance of the umbrella like plasma profile and streak picture and formation of bounded 3-dimensional plasma structures embedded within the pinch plasma without taking into account microscopic details of physical phenomena.

# I. Introduction:

A physical theory of plasma, by its very nature, is approximate, unlike say, optics, electromagnetism or mechanics. It is based on simplifying assumptions that assert that certain attributes of the plasma play a negligible role in governing some of its properties. A hierarchy of theories, which progressively relax a corresponding hierarchy of simplifying assumptions, is necessary to construct an 'asymptotically complete' description of plasma. The predictions of a true physical theory (for example, of optics or mechanics) must *necessarily* be realized in practice, which enables the theory to be the foundation of a technology. A partial theory, often referred as a model, does not possess this element of necessity and is therefore not mature enough to serve as the foundation of a technology. Working towards realization of a physical theory of a particular plasma, that actually is reliable enough to serve as a foundation of a technology, is therefore a worthwhile endeavor.

A kinematic model of plasma, which represents the shape and position of the plasma by an imaginary 3-dimensional surface (as a cartoon represents a human figure), may be considered as the ground level of such hierarchy of theories, which neglects almost all physical processes except perhaps some rudimentary conception of "motion" in a generalized sense. The Gratton-Vargas (GV) model [1] of the Dense Plasma Focus (DPF) [2,3,4] recently re-visited and revised [5-7] is an example of such a kinematic model. The present series of papers represents first steps towards an "asymptotically complete theory" of the Dense Plasma Focus that one day may hopefully enable the DPF to transition to a mature technology. It utilizes the basic approach of the GV model as its foundation.

The GV Model has demonstrated its ability [8] to fit experimental current waveforms from multiple installations, its utility as a scaling theory of device optimization [9,10] and its potential for providing a framework for constructing first-principles analyses of the DPF physics based on Hyperbolic Conservation Law (HCL) formalism [11,12,13]. It, however, continues to suffer from limitations: it fails to take into account the finite time interval between the start of current and plasma propagation [14], and the finite thickness of the dense plasma sheath [15], which governs the volume of the hot plasma in the pinch phase. This paper aims to construct a kinematic framework, very similar in its approach and spirit to the GV model but quite different in scope, that attempts mitigation of these limitations and in the process, also extends it to some modifications of the standard Mather type device geometry.

A unique aspect of the GV model is its kinematic and analytical nature. The GV model derives its results from mathematical properties of the analytical solution to a partial differential equation resulting from an apparently ill-founded hypothesis that resembles conservation of momentum. This is a departure from the usual practice of constructing models that use well-established fundamental principles of physics such as laws of mechanics and electrodynamics or physical phenomena [16] such as heat transport, electrical resistivity, atomic and molecular processes involving collisions / excitation / dissociation / ionization / radiation etc. There is no reason, rooted in physics, for this kind of model to yield results that compare well with experiments. That this model still manages to reproduce current waveforms [8] and plasma shapes similar to those experimentally determined [17] is itself a significant physical datum, that retrospectively justifies the assumptions made while deriving the partial differential equation.

The GV model is based on the *a priori* assumption of equality between the magnetic pressure due to the azimuthal magnetic field behind the plasma current sheath (PCS) and the mass density of the fill gas times the square of the normal component of sheath velocity. This could be "justified" as a boundary condition on the propagating PCS. Sufficiently far behind it, the magnetic pressure is the only force acting on a unit area of the PCS and sufficiently far ahead, the neutral gas of mass density $\rho_0$ getting ingested by the PCS moving with a normal velocity $v_{PCS}$ brings in a momentum density $-\rho_0 v_{PCS}$ with a momentum density flux $(-\rho_0 v_{PCS})(-v_{PCS})$. Newton's second law then suggests equality of the force per unit area with the momentum density flux. *This however is not a rigorous application of Newton's second law* (which must be applied at every point) and must be treated as a model assumption. This model assumption happens to be supported by the 2-D MHD simulations of Maxon and Eddleman [18] and experimental data [17].

This equality helps in defining a dimensionless time-equivalent parameter $\tau$, which equals the charge flow in real time t normalized to a 'mechanical equivalent of charge' [1] – a quantity having dimensions of charge constructed from physical parameters of the DPF. This equality also enables the GV model to construct a partial differential equation for the time and space evolution of an imaginary 3-dimensional surface of revolution – called GV surface – that is the anchor of this equality. It turns out that this family of 3-D surfaces resembles the experimentally observed shape of the PCS [17].

Assuming that this family of GV surfaces approximates the actual path of current flow, its circuit-element representation (usually identified as a time-varying inductance) can be calculated as a function of the dimensionless time-equivalent parameter $\tau$. The circuit equation then provides the current profile as a function of $\tau$, from which, the corresponding real time t can be determined as the *last procedure* of the model. *This ability to decouple the circuit equation from the physical motion of the plasma in real time is a feature that distinguishes the GV model from all other theories and models of DPF*. The original work of Gratton and Vargas [1] was unable to take into account the circuit resistance, without which, it was not possible to compare the model with experimental current waveforms. This was accomplished more recently [5] leading to the Resistive Gratton-Vargas (RGV) model.

It has been emphasized above that this is a kinematic model *which contains very little physics*. In particular, it does not contain conservation of mass and energy and various transport phenomena and its connection with conservation of momentum cannot be rigorously justified. It still manages to fit the current waveform of some DPF devices [8] using the static inductance, resistance and gas fill pressure as fitting parameters. This fitting process, however, leads to values of pressure and inductance that significantly deviate from those measured [4]. This deficiency indicates that there might be some scope for improving the kinematic model. At the same time, recent work [12,13] suggests that an improved kinematic model could be used as a base to construct a physical theory using the Hyperbolic Conservation Law (HCL) formalism.

The present Part-I of a planned series of papers attempts construction of a kinematic framework for describing some aspects of DPF that incorporates many features of the RGV model and serves as the foundation of a physical theory of the DPF introduced in Part-II. It differs from the GV model in four respects.

A. First, the present paper notes that the equations of motion of a plasma can be rendered into a dimensionless form for a dense plasma focus problem using time and space dependent scaling parameters related to the plasma focus device and the power source. The dimensionless equations in scaled variables then have no connection with the machine. This reduces the plasma focus problem into two sub-problems: (a) the evolution of the scaling parameters in real time and space (b) the evolution of scaled parameters in the non-dimensional time and space. The first sub-problem turns out to be exactly the kinematic framework of the GV model, eliminating the need for any ill-justified founding hypothesis. The second problem results in the time evolution of the distribution of plasma parameters in space – in other words, a physical theory. The physics content of this kinematic framework resides in the scaling properties of the equations of motion of the plasma in its single fluid approximation. This paper demonstrates that many aspects of the plasma focus phenomenology are mimicked by this theory.

B. Second, while the GV model is purely kinematic, the present framework borrows some insights from physical models, theories and experimental data external to the GV model. For example, the traditional representation of the plasma focus as a time varying inductance leads to conceptual difficulties in the interpretation of some experimental data [19]. These difficulties can be mitigated by a first-principles approach based on Poynting's Theorem [19], which requires calculation of quantities related to a circuit element representation from the shape and location of the current carrying layer, which is provided by the GV approach. Instead of a mere "cartoon representing a human figure ", this could be compared with a "pencil sketch with shading". The traditional representation of plasma focus as a time-varying inductance is shown to be a consequence of neglecting certain physical effects, which may appear as anomalous phenomena.

C. Third, it aims to incorporate in its kinematic structure, two experimentally observed features that are absent in the GV model. One is the time interval between the start of current and the start of PCS propagation [14]. The second is the presence of a finite sheath thickness [15,20].

D. Fourth, while GV model has been discussed so far only in the context of a Mather type plasma focus, the new effort attempts to incorporate some modifications of the standard Mather geometry.

The motivation for extending the discussion to non-standard adaptations of the simple Mather geometry arises from the fact that many laboratories do work with anodes which are tubular or have a cavity or protrusion near the axis or have a filleted, hemispherical or spherical anode. Presently, there is no formal theoretical framework for comparing the effect of such deviations on the operation taking the simple geometry as a reference. Although such framework is built into the new model, its emphasis continues to be on the understanding of the phenomenology of the classical Mather geometry simply because much still remains to be understood about it. Constructing this framework involves exploration of the mathematical properties of solutions of the GV partial differential equation and associated supplementary conditions purely within the context of a generalized plasma focus problem, which is going to remain an amateur effort until trained mathematicians take an interest.

In order to clearly establish the relation between this model and the GV model, the present paper revisits the construction of the GV model to give a concise and coherent review of the work referred above in a single logically consistent narrative and along the way shows how the missing aspects are being addressed in the new model. While earlier work [5,6], was more in the nature of tutorial introduction to the nearly forgotten work of Gratton and Vargas [1] after a gap of several decades (and also constrained by page limits of a conventional printed journal publication), the emphasis in this paper is on a comprehensive presentation of the GV model (including some missing details and correction of inconsistencies of nomenclature) and its elevation to an incrementally more descriptive theory. The treatment is more tutorial than is customary in scientific literature for the benefit of newcomers to this field.

This work is being presented in a series of papers with each part having a dominant theme. The focus of this Part I is on establishing the kinematic framework of the model including propagation delay and finite sheath thickness along with a clear statement of model assumptions. Future parts shall refer to results reported in Part I and focus on aspects other than purely kinematic (physical, numerical, applications etc.).

This paper is organized as follows. Section II derives a generalization of the RGV model that is valid for arbitrary algebraically defined shapes of anode and insulator and takes into account the sheath propagation delay. An illustration of the method for complex anode and insulator shapes is presented in Section III. Section IV introduces a finite sheath thickness as a kinematic artifact and explores its consequences. Section V discusses model closure. Section VI provides a summary of the main results and concludes the paper.

## II. Generalization of RGV model including modified geometry, sheath propagation delay

The RGV model [5,6] assumes that the DPF consists of a straight cylindrical anode of radius 'a' and length $\mathbb{L}_a$, an insulator of outer radius $\mathbb{R}_I$ and length $\mathbb{L}_I$ and a cathode of inner radius $\mathbb{R}_C$. It is filled with a gas having mass density $\rho_0$ and is powered by a capacitor bank that is idealized by a single capacitance $C_0$, in series with a constant resistance $R_0$, a constant inductance $L_0$ and an ideal closing switch (including its connection hardware), whose contributions to circuit inductance, circuit resistance and circuit capacitance are assumed to be included in $L_0$, $R_0$ and $C_0$. The capacitor bank parameters are defined to be external to the plasma focus device.

*This idealization of circuit may not be valid in some DPF devices.* In some small devices, the switch resistance may have significant variation as compared with the circuit impedance over a significant portion of the quarter cycle time. In some large devices, multiple spark gaps may begin conduction at significantly different times. In both cases, the assumed circuit model is not expected to be valid. This needs to be kept in mind when a fitting procedure fails to yield a good fit to the experimental current waveform with the RGV model or the present theory.

In this section, this description of the DPF is relaxed to include arbitrary shapes of anode and insulator, which are represented by surfaces of revolution about the z-axis in cylindrical coordinates. *Not all shapes of anode and insulator are amenable to this treatment – the limitations are discussed later in the paper*. For this purpose, 'a' is re-defined as a normalizing scale length related to the anode size. The anode and insulator shapes are defined in terms of a coordinate $\tilde{Z}$ (which is used only for specifying boundaries) measured in units of scale length 'a' along the cylindrical axis starting from the base of the anode. In a two-dimensional representation of these 3-D surfaces of rotation in $(r,z)$ space, the anode would be a piece-wise continuous curve that has a finite radius at its base $\tilde{Z}=0$ and would have a maximum height $\tilde{Z}=\tilde{z}_A$, the normalized length of the anode. Similarly, the insulator would be a curve that begins at $\tilde{Z}=0$ with a finite radius and *meets the anode* at $\tilde{Z}=\tilde{z}_I$, the normalized length of the insulator. The normalized anode radius $\tilde{R}_A$ (more accurately, the radial coordinate of a point on the anode surface whose axial coordinate equals $\tilde{Z}$) and normalized insulator radius $\tilde{R}_I$ (similar to the case of anode radius) are specified as functions $\tilde{R}_A(\tilde{Z})$ and $\tilde{R}_I(\tilde{Z})$ of $\tilde{Z}$, which may be piecewise continuous with arbitrary number of segments:

$$\begin{aligned}\tilde{R}_A(\tilde{Z}) &= \tilde{R}_{A,ext}^{(i)}(\tilde{Z}), \quad \tilde{Z}_A^{(i-1)} \leq \tilde{Z} \leq \tilde{Z}_A^{(i)}, i = 1,2\cdots,i_m; \quad \tilde{R}_{A,ext}^{(i)}(\tilde{Z}_A^{(i)}) = \tilde{R}_{A,ext}^{(i-1)}(\tilde{Z}_A^{(i)}); \\ &= \tilde{R}_{A,int}^{(j)}(\tilde{Z}), \quad \tilde{Z}_A^{(i_m+j-1)} \geq \tilde{Z} \geq \tilde{Z}_A^{(i_m+j)}, j = 1,2\cdots,j_m; \quad \tilde{R}_{A,int}^{(j)}(\tilde{Z}_A^{(i_m+j)}) = \tilde{R}_{A,int}^{(j+1)}(\tilde{Z}_A^{(i_m+j)});\end{aligned}$$ (1)

$$\tilde{R}_I(\tilde{Z}) = \tilde{R}_I^{(i)}(\tilde{Z}), \quad \tilde{Z}_I^{(i-1)} \leq \tilde{Z} \leq \tilde{Z}_I^{(i)}, i = 1,2\cdots,i_n; \quad \tilde{R}_I^{(i)}(\tilde{Z}_I^{(i)}) = \tilde{R}_I^{(i-1)}(\tilde{Z}_I^{(i)});$$ (2)

The anode profile consists of two branches: an external branch consisting of segments $\tilde{R}_{A,ext}^{(i)}(\tilde{Z})$, where the coordinate $\tilde{Z}$ increases monotonically till it reaches a maximum value $\tilde{Z}_A^{(i_m)}$, which is designated as the normalized anode length $\tilde{z}_A$. After that, the anode profile may or may not have an internal branch representing a cavity, consisting of segments $\tilde{R}_{A,int}^{(j)}(\tilde{Z})$, where $\tilde{Z}$ would decrease monotonically. The anode profile segment functions $\tilde{R}_{A,ext}^{(i)}(\tilde{Z})$ and $\tilde{R}_{A,int}^{(j)}(\tilde{Z})$ will not henceforth be distinguished in the text unless the context requires such distinction and both will be referred as $\tilde{R}_A^{(i)}(\tilde{Z})$. The points $(\tilde{R}_A^{(i)}(\tilde{Z}_A^{(i)}), \tilde{Z}_A^{(i)})$ etc. shall be referred as vertices. The following convention is also adopted: $\tilde{Z}_A^{(0)} \equiv \tilde{z}_I$, the normalized length of the insulator, and $\tilde{Z}_I^{(0)} = 0$.

The cathode is still taken to be a hollow cylinder with inner normalized radius $\tilde{r}_C$ connected to a metal base plate lying in the $\tilde{Z} = 0$ plane. Illustration of the present model with simple multi-segment anode profiles, including a real Mather type plasma focus, is discussed in Section III.

The power source is still assumed to be a capacitor bank as in the case of the RGV model. However, this is merely for the sake of convenience in the subsequent discussion. It will be made clear at various stages of the discussion that this assumption can be relaxed in an appropriate manner.

The subsequent discussion is partitioned into three subsections. The first deals with the physical foundations of the kinematic model. The second deals with the phase between the start of current and the start of plasma propagation. The third deals with the propagation of the plasma.

### A. Physical foundations of the kinematic model:

As discussed in the Introduction, the RGV model is based on the *a priori* assumption of equality between the magnetic pressure due to the azimuthal magnetic field behind the plasma current sheath (PCS) and the mass density of the fill gas times the square of the normal component of sheath velocity, which leads to certain mathematical consequences. This subsection looks at an alternate approach to arriving at the same mathematical results starting with more fundamental physical principles.

This approach looks for the conditions under which the equation of motion for the plasma

$$\rho\left\{\frac{\partial \vec{v}}{\partial t} + (\vec{v} \cdot \vec{\nabla})\vec{v}\right\} = \vec{J} \times \vec{B} - \vec{\nabla}p \tag{3}$$

along with Ampere's Law

$$\vec{\nabla} \times \vec{B} = \mu_0 \vec{J} \tag{4}$$

can be written in a dimensionless form for the plasma focus problem. Clearly, the local mass density $\rho$ must be scaled with the initial fill gas density $\rho_0$, the coordinates must be scaled with the scale length 'a' of the plasma focus anode and the magnetic field must be scaled with the azimuthal magnetic field produced by the instantaneous current $I(t)$ at radius r. Denoting the dimensionless quantities with an overtilde and scaling parameters by the subscript 0, (for example, $\rho = \rho_0 \tilde{\rho}$; $\vec{v} = v_0 \tilde{v}$; $r = a\tilde{r}$; $z = a\tilde{z}$; $\vec{\nabla} = a^{-1}\tilde{\nabla}$; $\vec{B} = B_0 \tilde{B}$; $B_0 \equiv B_{00}(t)\tilde{r}^{-1}$; $B_{00}(t) \equiv \mu_0 I(t)/2\pi a$; $p = p_0 \tilde{p}$)

(4) can be written as

$$\vec{J} = (\mu_0 a)^{-1} B_{00}(t) \tilde{\nabla} \times (\tilde{r}^{-1}\tilde{B}) \tag{5}$$

and (3) can be written as

$$\rho_0 \tilde{\rho}\left\{\frac{\partial(v_0 \tilde{v})}{\partial t} + a^{-1}v_0(\tilde{v} \cdot \tilde{\nabla})v_0 \tilde{v}\right\}$$
$$= B_{00}^2(t)(\mu_0 a)^{-1} \tilde{r}^{-1}(\tilde{\nabla} \times (\tilde{r}^{-1}\tilde{B})) \times \tilde{B} - a^{-1}\tilde{\nabla}p_0 \tilde{p} \tag{6}$$

Define

$$v_0 \equiv \frac{B_0}{\sqrt{2\mu_0\rho_0}} = \frac{B_{00}(t)}{\sqrt{2\mu_0\rho_0}} \tilde{r}^{-1} = \frac{\mu_0 I(t)}{2\pi a \sqrt{2\mu_0\rho_0}} \tilde{r}^{-1}; \tag{7}$$

$$\frac{d\tau}{dt} = \frac{I(t)}{Q_m} \Rightarrow \frac{\partial}{\partial t} \equiv \frac{d\tau}{dt}\frac{\partial}{\partial \tau} = \frac{I(t)}{Q_m}\frac{\partial}{\partial \tau} \tag{8}$$

$$p_0 \equiv \frac{B_0^2}{2\mu_0} = \frac{B_{00}(t)^2}{2\mu_0}\tilde{r}^{-2} = \frac{\mu_0 I^2(t)}{8\pi^2 a^2 \tilde{r}^2} \tag{9}$$

It can now be seen that the equation of motion (3) can be made dimensionless by choosing

$$Q_m = \pi\mu_0^{-1} a^2 \sqrt{2\mu_0\rho_0} ; \tag{10}$$

and can be written as

$$\tilde{\rho}\left\{\frac{\partial \tilde{v}}{\partial \tau} + \tilde{v}\frac{\partial}{\partial \tau}\text{Log}(I(\tau)) + \frac{1}{2}\tilde{v}\cdot\tilde{\nabla}\left(\tilde{r}^{-1}\tilde{v}\right)\right\} \\ = -\tilde{B}\times\left(\tilde{\nabla}\times\left(\tilde{r}^{-1}\tilde{B}\right)\right) - \frac{1}{2}\tilde{r}\tilde{\nabla}\left(\tilde{r}^{-2}\tilde{p}\right) \tag{11}$$

This non-dimensional form of the equation of motion is independent of the device parameters and the power source. It can be verified that the equations of conservation of mass and energy, the Generalized Ohm's Law and Maxwell's equations can also be rendered dimensionless using the same scaling. It will be seen during the course of this discussion that the use of space and time dependent scaling parameters enables separation of the physical problem into two sub-problems: (1) Evolution of the scaling parameters in space and time via the Gratton-Vargas equation (2) the spatial and temporal distribution of the scaled parameters: density $\tilde{\rho}$, velocity $\tilde{v}$, magnetic field $\tilde{B}$ and pressure $\tilde{p}$ in dimensionless space and time.

The first part dealing with evolution of the scaling parameters in space and time is the domain of the kinematic framework which is the subject of this paper. It is shown in Section II C that the scaling relations (7)-(10) are exactly equivalent to the fundamental assumption of the GV model. The second part leads to a physical theory, a topic for Part II.

**B. The phase between the start of current and start of plasma propagation**:

The operation of the ideal switch connects the anode to the high voltage terminal of the capacitor bank, so that a potential difference appears across the insulator. A plasma layer is formed at the interface between the insulator and the neutral gas, whose conductivity increases from zero to a sufficiently high value for current conduction over a finite time, during which, the current is negligibly small. This interval between the application of voltage and start of current is known as the plasma formation phase [4], which is outside the scope of the present paper.

After the start of current, the weakly ionized and partially dissociated deuterium gas starts getting heated resistively. At the same time, the repulsive magnetic force between the current carrying anode and the plasma layer pushes the plasma away from the anode, effectively detaching from the insulator.

A model of this lift-off phase has been described by Bruzzone, Acuña and Clausse [14]. It calculates the time required for the plasma to move through a radial distance equal to its thickness, whose value is determined through an experiment. This work demonstrates that many physical phenomena need to be accounted for, many times through reasonable approximations or assumptions, in order to calculate the time interval between the first peak of the experimental current derivative waveform and observable displacement of the plasma. One of the conclusions of this work is that the lift-off is a supersonic process.

The problem of supersonic propagation of a magnetically driven plasma has been examined [11] using a Hyperbolic Conservation Law formulation, using equations for conservation of mass, momentum and energy in a frame of reference co-located with the sheath. The treatment is analogous to that of a detonation wave problem in an energetic material, which takes into account the specific energy released by exothermic processes behind the shock front in the energy conservation equation. In the case of the magnetically driven plasma, the role of the exothermic energy source is played by the excess of electromagnetic energy, resistively deposited behind the shock wave by the current supplied from an external power source via electrodes, over the energy consumed in dissociation and ionization of the fill gas.

This model makes the assumption of local planarity [21]: neglect of plasma curvature over dimensions larger than the thickness of the shock discontinuity. Equivalently, the order of gradient scale length in a direction normal to the sheath is assumed to be much smaller than the order of the radius of curvature. Effectively, this amounts to neglecting the spatial variation of $\tilde{r}^{-1}$ in equation (11). It also has a natural scale of the plasma velocity given by (7).

Its main finding is that there is a lower bound $v_{LB} \propto \sqrt{\varepsilon_{d+i}}$ on this scaling velocity $v_0(t,r)$ related to the specific energy $\varepsilon_{d+i} \sim 7.45 \times 10^8 \, J/kg$ for dissociation and ionization for deuterium. The supersonic propagation of the shock discontinuity can occur only when $v_0 \geq v_{LB}$.

The existence of this lower bound for supersonic propagation is a consequence of the requirement that the current carrying portion of the plasma must be "hot enough" to possess a "sufficient" electrical conductivity. The temperature behind the supersonic shock wave is related to the average ion kinetic energy, which is proportional to the square of the scaling velocity. This lower bound on the temperature behind the shock therefore translates into a lower bound $v_{LB} = f_{LB}\sqrt{\varepsilon_{d+i}}$ on the scaling velocity, where $\varepsilon_{d+i} \sim 7.45 \times 10^8 \, J/kg$ is the specific energy for dissociation and ionization for deuterium and $f_{LB}$ is a factor that needs to be empirically determined as an exact specification of how hot the plasma really needs to be to take off. The HCL analysis [11] assumes full dissociation and ionization as a criterion, necessary for formation of thin current layers associated with good pinch formation. Detachment from the insulator [14] can occur when the plasma is partially dissociated and ionized [14], requiring a much smaller threshold.

The time interval $t_L$ between the start of current and the start of supersonic plasma propagation can then be interpreted as the time when the following equality holds:

$$v_0(t_L, R_{ins}) = \frac{1}{\sqrt{2\mu_0\rho_0}} \frac{\mu_0 I(t_L)}{2\pi R_{ins}} = v_{LB} = f_{LB}\sqrt{\varepsilon_{d+i}} = 2.9 \times 10^4 f_{LB} \quad (12)$$

where $R_{ins}$ is the maximum radius of the insulator. This gives

$$I(t_L) = v_{LB} \frac{2\pi R_{ins}}{\mu_0} \sqrt{2\mu_0\rho_0} \quad (13)$$

During this period, the displacement of the plasma is negligible and the circuit inductance does not differ appreciably from the static inductance of the capacitor bank.

The current in the circuit then follows the standard waveform for a capacitor discharge circuit with an ideal switch and an initial charge $Q_0 = C_0 V_0$

$$I_s(t) = \frac{I_0}{\sqrt{1-\lambda^2}} \exp(-\gamma_0 t) \sin(\omega_0 t), \quad 0 \leq t \leq t_L \tag{14}$$

$$I_0 = V_0 \sqrt{\frac{C_0}{L_0}} \; ; \; \gamma_0 = \frac{R_0}{2L_0} \; ; \; \lambda \equiv \frac{R_0}{2}\sqrt{\frac{C_0}{L_0}} \; ; \; \omega_0 = \sqrt{\left(\frac{1}{L_0 C_0} - \gamma_0^2\right)} = \frac{1}{\sqrt{L_0 C_0}}\sqrt{(1-\lambda^2)} \; ; \tag{15}$$

The time $t_L$ can be determined from equation (12) rewritten as

$$\exp(-\gamma_0 t_L)\sin(\omega_0 t_L) = v_{LB}\sqrt{2\mu_0 \rho_0}\,\frac{2\pi R_{ins}}{\mu_0 I_0}\sqrt{1-\lambda^2} \; ; \tag{16}$$

Equation (16) has no solution for $t_L$ when

$$\rho_0 > \rho_{0,max}\,(kg/m^3) = \frac{1}{2\mu_0 v_{LB}^2}\left(\frac{\mu_0 I_{max}}{2\pi R_{ins}}\right)^2$$

$$= 18.83 \frac{1}{f_{LB}^2}\left(\frac{I_{max}(MA)}{R_{ins}(mm)}\right)^2 \text{ (for deuterium)} \tag{17}$$

where

$$I_{max} = \frac{I_0}{\sqrt{1-\lambda^2}} \text{Max}\left[\exp(-\gamma_0 t)\sin(\omega_0 t)\right] = I_0 \exp\left(-\frac{\lambda \text{Arc}\cos(\lambda)}{\sqrt{1-\lambda^2}}\right) \tag{18}$$

A plasma focus discharge at a high enough mass density according to (17) will therefore have a current waveform given by (14), which can be used to determine the constants $L_0$ and $R_0$ as well as the calibration constant for the current monitor.

The charge flow until $t_L$ for this circuit is

$$Q_L = Q_0 \left\{ 1 - \exp\left(-\frac{\lambda}{\sqrt{(1-\lambda^2)}} \omega_0 t_L \right) \left( \cos(\omega_0 t_L) + \frac{\lambda}{\sqrt{(1-\lambda^2)}} \sin(\omega_0 t_L) \right) \right\} \qquad (19)$$

### C. The plasma propagation phase:

After the scaling velocity $v_0$ crosses the lower velocity bound $v_{LB}$, the supersonic propagation of the plasma begins. The hydrodynamic shock wave becomes an ionizing shock wave [21] when the photons produced in the plasma travel ahead of the shock wave, get absorbed in the neutral gas layer adjacent to the shock wave and produce photoelectrons. The current carrying layer of plasma has an electric field [21] in a direction perpendicular to the direction of propagation of plasma that is of the order of $\vec{v} \times \vec{B}$ where $\vec{v}$ is the fluid velocity of the plasma and $\vec{B}$ is the magnetic field in the current carrying layer. Since the transverse component of electric field is continuous across interfaces between dissimilar media, this electric field extends into the neutral gas layer adjacent to the hydrodynamic shock wave, accelerates the photo-electrons, dissociates and ionizes the neutral gas layer by electron impact ionization and further heats it resistively [21]. As a result, the adjacent layer of neutral gas gets incorporated into the moving ionizing shock wave, which eventually evolves into an autonomous structure consisting of a fully ionized plasma zone that carries the bulk of the current and another transition zone [15] that contains a partially ionized plasma which is bounded on one side by neutral gas and on the other side by a fully ionized plasma, capable of propagating by itself into the neutral gas. This autonomous structure shall be referred as a plasma current sheath (PCS) and its transition zone between the neutral gas and the fully ionized plasma shall be called as the dense plasma sheath (DPS). This scenario is often referred as the snowplow model since it assumes that the neutral gas layer adjacent to the hydrodynamic shock wave gets fully incorporated into the PCS just as the snowplow gathers the entire layer of snow in its path.

Note that the above description implicitly assumes that the ambient geo-magnetic field and any dynamo processes that might amplify it do not significantly affect the evolution of the PCS. It does not contend that such effects are completely absent for all time. This situation would however change if a sufficiently strong poloidal magnetic field is externally imposed which could get amplified by dynamo processes and become comparable with the azimuthal magnetic field. Such an experimental condition would negate the foundation of the present theory.

The propagating PCS has a scale of velocity $v_0(t,r)$ given by (7). One can now define an imaginary surface, whose local normal velocity is given by

$$v_n = \chi v_0 = \chi \tilde{r}^{-1} \frac{B_{00}(t)}{\sqrt{2\mu_0 \rho_0}} ; \chi \sim 1 \qquad (20)$$

At this surface, the equality

$$\rho_0 v_n^2 = \chi^2 B_0^2 / 2\mu_0 ; \qquad (21)$$

holds *by definition*. This may be loosely considered as a statement of global momentum balance in a heuristic manner, as described in the Introduction as the basis of the RGV model. *Note, however, that this isn't local momentum balance implied by Newton's Second Law: the neutral gas and the magnetic field exist in mutually exclusive regions* [22].

Here, $\chi$ is a number which equals 1 for the GV model and is meant as a placeholder for its generalization. Let this surface, called the GV surface, having full azimuthal symmetry, be described in cylindrical coordinates by the equation

$$\psi(r,z,t) \equiv z - f(r,t) = 0 \qquad (22)$$

By definition, the convective derivative of $\psi$ is zero:

$$\partial_t \psi + \vec{v} \cdot \vec{\nabla} \psi = 0 \qquad (23)$$

where $\vec{v}$ is the velocity of the surface. The component of velocity normal to the surface is

$$v_n = \vec{v} \cdot (\vec{\nabla}\psi / |\vec{\nabla}\psi|) = -\partial_t \psi / \sqrt{(\partial_r \psi)^2 + (\partial_z \psi)^2} \qquad (24)$$

From (21) and (24), using the standard result from Ampere's Law

$$B_0 = \frac{\mu_0 I(t)}{2\pi r} \qquad (25)$$

used for the scaling scheme, one can write

$$\partial_t \psi + \sqrt{(\partial_r \psi)^2 + (\partial_z \psi)^2} \frac{\chi \mu_0 I(t)}{2\pi r \sqrt{2\rho_0 \mu_0}} = 0 \qquad (26)$$

*Note that this result applies only when the PCS structure has been formed and has started its magnetically driven autonomous propagation, i.e., after the time $t_L$, in view of the analysis of the electromagnetic analog of the detonation wave problem [11].*

Define now a time-equivalent dimensionless parameter $\tau$ (which shall often be called 'time' for convenience and the time t will be referred as 'real time') by the equation [1,5]

$$\tau \equiv \frac{\chi}{Q_m} \int_{t_L}^{t} I(t)\,dt \quad \text{where} \quad Q_m = \mu_0^{-1} \pi a^2 \sqrt{2\mu_0 \rho_0}\,, \qquad (27)$$

and also define dimensionless coordinates $\tilde{r} \equiv r/a$, $\tilde{z} \equiv z/a$. This enables reduction of (26) to a dimensionless form

$$\partial_\tau \psi + \sqrt{(\partial_{\tilde{r}} \psi)^2 + (\partial_{\tilde{z}} \psi)^2} \frac{1}{2\tilde{r}} = 0 \qquad (28)$$

This is the Gratton-Vargas [1,5] equation, *where the definition of $\tau$ has now been altered to include the time interval between start of current and start of sheath propagation and the generalization constant $\chi$*. Note that the parameter 'a', which signified anode radius in previous work [5-7], has been re-defined as a scale length related to the generalized anode shape, since the anode radius varies along its length.

This equation is based on precisely four inputs:

(a) definition of normal velocity of an arbitrary propagating surface of revolution in cylindrical geometry (equation (24))

(b) the definition of the normal velocity of the imaginary surface in terms of the scale velocity (equation (20))

(c) a standard result from Ampere's Law (equation (25)) and

(d) definition of the variable $\tau$ (equation (27)).

*It specifically does not involve the following*:

(1) the assumption of a specific shape or size of anode, cathode or insulator or any relations between them *and indeed even their existence*, (see the end of Section IIIa)

(2) the precise definition of the scale length 'a' which is *not* the anode radius as in earlier work [5,6],

(3) the assumption that the driving circuit is a simple capacitor discharge circuit with constant intrinsic inductance and resistance.

The Gratton-Vargas equation can now be seen to be a direct result of the use of space and time dependent scaling parameters in making the equation of motion dimensionless.

A crucial aspect of the GV model is the dual nature of the variable $\tau$. In the GV partial differential equation (28), along with applicable supplementary conditions discussed below, it is clearly a geometrical variable, having no relation with the power source and its electrical characteristics. It has a relation with the scaled anode and insulator profile, as will become clear shortly (see the derivation of equation (50)). On the other hand, by its definition (27), it clearly has no relation with the scaled geometry and is related solely with the operating parameters of the device and the driving circuit. This dual nature of $\tau$ enables the GV model and its adaptations to decouple the plasma propagation from the driving circuit. For example, regardless of whether the device is powered by a capacitor bank or an inductive energy storage, the propagation of the GV surface in the dimensionless $(\tilde{r},\tilde{z})$ space as a function of $\tau$ remains absolutely unchanged. What changes is the recovery of the real time t from $\tau$ using the expression for dimensionless current as function of $\tau$ after solving the circuit equation for the particular kind of energy storage system.

For its application to the generalized plasma focus problem stated above, equation (28) must be considered along with the following supplementary conditions.

1. At $\tau = 0$, the initial shape of the solution corresponds to the insulator profile, on which the initial plasma is formed: $\psi(\tilde{r},\tilde{z},\tau = 0) = \psi(\tilde{R}_I(\tilde{Z}),\tilde{Z},\tau = 0)$

2. At every value of $\tau > 0$, the solution has a curve of intersection with the anode; i.e., $\psi(\tilde{R}_A(\tilde{Z}),\tilde{Z},\tau)$ is part of the solution of the equation at $\tau$.

3. It is assumed that the current that crosses over from the anode to the plasma does so *at or behind the GV surface*. That means that the current density at the anode surface has a zero in-surface component at the junction between the GV surface and the anode. Assuming further (in view of the similarity of experimentally determined sheath shape with the solutions of the GV equation [17]), that the GV surface approximates the shape of the path that the current density takes in the plasma, the condition of continuity of current density at the junction implies that *the GV surface must be perpendicular to the anode at the junction*. This condition applies only to the forward motion of the GV surface and not to the reflected motion that begins after the plasma reaches the center. It also does not apply to the motion of the GV surface in the internal branch of the anode profile.
4. The solution must join the anode profile with the cathode.

The full extent of mathematical properties of this equation, the supplementary conditions and its solutions has not been explored till date and many uncertainties and doubts remain about the existence, uniqueness and character of its solution for various kinds of boundaries. A tutorial discussion on some types of solution of this equation given by Gratton and Vargas is found elsewhere [1,5,6]. Only a few salient features of the solution [1,5] using the method of characteristics [23] are brought out here. The discussion is tailored for an audience of physicists with working knowledge of mathematics and not for professional mathematicians who are likely to be disappointed with the rudimentary mathematical rigor and might find this a fit topic for their expert attention.

The method of characteristics finds [23] a family of characteristic line elements (CLE) *in the neighborhood* of every point on the integral surface of the partial differential equation (1) along which its solution is constant and (2) which is locally perpendicular to the integral surface. A collection of such CLEs is referred as "characteristic curves". Since the integral surface moves with time, the characteristic line elements in its neighborhood are implicitly tagged with time, although the algebraic definition of the characteristic curve may not have time as a variable. This subtle distinction has important consequences which are discussed later.

Define generalized momenta $p_{\tilde{r}} \equiv \partial_{\tilde{r}}\psi$, $p_{\tilde{z}} \equiv \partial_{\tilde{z}}\psi$ and Hamiltonian $H = (2\tilde{r})^{-1}\sqrt{p_{\tilde{r}}^2 + p_{\tilde{z}}^2}$. Then (28) takes the Hamilton-Jacobi form $p_\tau + H = 0$ for $p_\tau \equiv \partial_\tau \psi$. The Hamiltonian equations for the characteristic line elements are then

$$\frac{d\tilde{r}}{d\tau} = \frac{\partial H}{\partial p_{\tilde{r}}} = \frac{p_{\tilde{r}}}{2\tilde{r}\sqrt{p_{\tilde{r}}^2 + p_{\tilde{z}}^2}} = \frac{p_{\tilde{r}}}{4\tilde{r}^2 H} \qquad (29)$$

$$\frac{d\tilde{z}}{d\tau} = \frac{\partial H}{\partial p_{\tilde{z}}} = \frac{p_{\tilde{z}}}{2\tilde{r}\sqrt{p_{\tilde{r}}^2 + p_{\tilde{z}}^2}} = \frac{p_{\tilde{z}}}{4\tilde{r}^2 H} \qquad (30)$$

$$\frac{dp_{\tilde{r}}}{d\tau} = -\frac{\partial H}{\partial \tilde{r}} = \sqrt{p_{\tilde{r}}^2 + p_{\tilde{z}}^2}\,\frac{1}{2\tilde{r}^2} = \frac{H}{\tilde{r}} \qquad (31)$$

$$\frac{dp_{\tilde{z}}}{d\tau} = -\frac{\partial H}{\partial \tilde{z}} = 0 \qquad (32)$$

This system of equations has two invariants. Equation (32) shows $p_{\tilde{z}}$ to be one; H can be shown to be the other [1].

$$\frac{dH}{d\tau} = \underbrace{\frac{\partial H}{\partial \tau}}_{=0} + \frac{\partial H}{\partial p_{\tilde{r}}}\frac{dp_{\tilde{r}}}{d\tau} + \underbrace{\frac{\partial H}{\partial p_{\tilde{z}}}\frac{dp_{\tilde{z}}}{d\tau}}_{=0} + \frac{\partial H}{\partial \tilde{r}}\frac{d\tilde{r}}{d\tau} + \underbrace{\frac{\partial H}{\partial \tilde{z}}\frac{d\tilde{z}}{d\tau}}_{=0} = -\frac{\partial H}{\partial p_{\tilde{r}}}\frac{\partial H}{\partial \tilde{r}} + \frac{\partial H}{\partial \tilde{r}}\frac{\partial H}{\partial p_{\tilde{r}}} = 0 \qquad (33)$$

Combine the two invariants to define a new invariant N:

$$N \equiv \frac{p_{\tilde{z}}}{2H} \qquad (34)$$

The definition of H then gives [1]

$$p_{\tilde{z}} = \pm \frac{Np_{\tilde{r}}}{\sqrt{(\tilde{r}^2 - N^2)}}\,; \qquad (35)$$

Equations (29) and (30) then give

$$\frac{d\tilde{z}}{d\tau} = \frac{p_{\tilde{z}}}{4\tilde{r}^2 H} = \frac{N}{2\tilde{r}^2} \tag{36}$$

$$\frac{d\tilde{r}}{d\tau} = \frac{p_{\tilde{r}}}{4\tilde{r}^2 H} = \frac{s}{2\tilde{r}^2}\sqrt{\left(\tilde{r}^2 - N^2\right)} \ ; \ s = \pm 1 \tag{37}$$

Combining (36) and (37), the cosine of the angle $\phi$ made by the local normal to the GV surface with the z axis is given by

$$\cos\phi = \frac{d\tilde{z}/d\tau}{\sqrt{\left(d\tilde{z}/d\tau\right)^2 + \left(d\tilde{r}/d\tau\right)^2}} = \tilde{r}^{-1}N \ ; \ \sin\phi = \pm\sqrt{1-\cos^2\phi} = \pm\tilde{r}^{-1}\sqrt{\left(\tilde{r}^2 - N^2\right)} \tag{38}$$

Equation (38) provides a geometrical interpretation to the quantity N. In terms of the normal velocity $v_n$ given by (21), the axial and normal components of velocity are given by

$$v_z = v_n \cos\phi, \ v_r = \pm v_n \sin\phi \tag{39}$$

It also shows that N is always less than $\tilde{r}$ and N is positive when $\phi$ is either in the first or the 4th quadrant. When the solution travels in the negative $\tilde{z}$ direction, for example, inside a cavity in the anode represented by the internal branch of the anode profile, N is clearly negative. The equation of the CLE is obtained from

$$\frac{d\tilde{z}}{d\tilde{r}} = \frac{d\tilde{z}}{d\tau}\bigg/\frac{d\tilde{r}}{d\tau} = s\frac{N}{\sqrt{\tilde{r}^2 - N^2}} \tag{40}$$

For $N \neq 0$, integration of (40) gives the relation describing the family $\mathbb{C}$ of curves of constant N and H, which are perpendicular to the GV surface[1,5]:

$$\frac{\tilde{z}}{N} - s\text{ArcCosh}\left(\frac{\tilde{r}}{|N|}\right) = C_1 = \text{constant} \qquad s = \pm 1 \tag{41}$$

Note that (37) is invariant under the transformation $N \to -N$ while (41) is invariant under the combined operation $N \to -N, \tilde{z} \to -\tilde{z}$. Integration of (37) gives the location of the integral surface on the family $\mathbb{C}$ of curves at time $\tau$.

$$\frac{\tilde{r}}{|N|}\sqrt{\frac{\tilde{r}^2}{N^2}-1}+\text{ArcCosh}\left(\frac{\tilde{r}}{|N|}\right)-\frac{s\tau}{N^2}=C_2=\text{constant} \qquad (42)$$

For the case of N=0, (37) shows

$$\tilde{r}^2 = s\tau + C_3, \qquad (43)$$

The constants of integration $C_1, C_2, C_3$ are to be determined by the specifications of the problem as discussed below.

In addition to the relations (41), (42) and (43), the solution must satisfy supplementary conditions mentioned above. The nature of the characteristic curves and GV surface for the case of Mather type plasma focus has been discussed in detail in Ref. 5. Note that equations (41) and (42) do not admit any scaling other than the normalization of the coordinates by the scale length 'a', which, however, is completely arbitrary.

The value associated with the symbolic sign s is determined for each segment or vertex from the physical context: it must satisfy the supplementary conditions in all segments and at all vertices. In this paper, the sign s is symbolically defined such that $s \Leftrightarrow \text{sign}(v_r)$ to conform with (37). Note that this symbolic definition of s is different from previously published work and needs to be taken into account when comparing results of this paper with earlier work. In general, there would be a unique choice of s for each segment and vertex that satisfies the supplementary conditions for a given profile. The assignment of the sign s needs to be graphically checked in practice as illustrated in Section III.

The first supplementary condition mentioned above implies that the characteristic should be perpendicular to the given insulator profile at its intersection. From (40),

$$\frac{d\tilde{R}_I(\tilde{Z})}{d\tilde{Z}} = -\frac{d\tilde{z}}{d\tilde{r}} = -s\frac{N_I(\tilde{Z})}{\sqrt{\tilde{R}_I^2(\tilde{Z})-N_I^2(\tilde{Z})}} \qquad (44)$$

Therefore, in the insulator (or plasma formation) region

$$N_I(\tilde{Z}) = \frac{\tilde{R}_I(\tilde{Z})|d\tilde{R}_I(\tilde{Z})/d\tilde{Z}|}{\sqrt{(d\tilde{R}_I(\tilde{Z})/d\tilde{Z})^2 + 1}} \tag{45}$$

Similarly, the third supplementary condition implies that the characteristic should be tangent to the anode profile at its intersection. Using (40)

$$\frac{d\tilde{r}}{d\tilde{z}} = s\frac{\sqrt{\tilde{r}^2 - N_A^2(\tilde{Z})}}{N_A(\tilde{Z})} = \frac{d\tilde{R}_A(\tilde{Z})}{d\tilde{Z}} \tag{46}$$

This gives

$$N_A(\tilde{Z}) = S\frac{\tilde{R}_A(\tilde{Z})}{\sqrt{1 + (d\tilde{R}_A/d\tilde{Z})^2}}; S = \pm 1 \tag{47}$$

The second supplementary condition implies that the point of intersection between the GV surface and anode profile in the $(\tilde{r},\tilde{z})$ plane must move along the anode profile as a function of $\tau$ as described by (36), where $\tilde{r} = \tilde{R}_A(\tilde{Z})$ and $\tilde{z} = \tilde{Z}$:

$$\frac{d\tilde{Z}}{d\tau} = \frac{N_A(\tilde{Z})}{2\tilde{R}_A^2(\tilde{Z})} \tag{48}$$

From (48), it is clear that the plus sign must be chosen in (47) when the solution is expected to begin at $\tilde{Z} = 0$ and move towards higher $\tilde{Z}$ as in the case of the external branch of anode profile. There may be situations such as a hollow anode or a cavity in the anode face, (represented by the internal branch in anode profile function (1)) when the solution is expected to move within the cavity towards decreasing values of $\tilde{Z}$. In such case, the negative sign would apply in (47). One could then set $S = \text{Sign}[d\tilde{Z}/d\tau]$.

Integration of (48) using (47) gives the value of $\tau(\tilde{Z})$ when the GV surface reaches the point $P_A(\tilde{R}_A(\tilde{Z}), \tilde{Z})$ on the external branch of anode profile

$$\tau(\tilde{Z}) - \tau_{i-1} = 2S \int_{\tilde{Z}_A^{(i-1)}}^{\tilde{Z}} \sqrt{1 + \left(d\tilde{R}_A^{(i)}/d\tilde{Z}\right)^2} \tilde{R}_A^{(i)}(\tilde{Z}) d\tilde{Z}; \quad \tilde{Z}_A^{(i-1)} \leq \tilde{Z}(\tau) \leq \tilde{Z}_A^{(i)}; \tau_0 = 0 \qquad (49)$$

The dimensionless time $\tau_i$ at which the GV surface reaches the vertex $\tilde{Z}_A^{(i)}$ is given by the recursion equations

$$\tau_i - \tau_{i-1} = 2S \int_{\tilde{Z}_A^{(i-1)}}^{\tilde{Z}_A^{(i)}} \sqrt{1 + \left(d\tilde{R}_A^{(i)}/d\tilde{Z}\right)^2} \tilde{R}_A^{(i)}(\tilde{Z}) d\tilde{Z} \qquad (50)$$

where $\tau_i - \tau_{i-1}$ is the $i^{th}$ time zone that corresponds to the $i^{th}$ segment of the anode profile.

The function $\tau(\tilde{Z})$ may be numerically inverted to obtain the coordinates of the intersection of GV surface with the external anode surface for a given $\tau$. For the RGV model case applicable to Mather type plasma focus, (50) reproduces the expression for the rundown time as $\tau_R = 2(\tilde{z}_A - \tilde{z}_I)$

It is important to note that the function $\tau(\tilde{Z})$ is a monotonically increasing function within each segment of external branch of anode profile. It implies that characteristic curves of family $\mathbb{C}$ originating on the external branch of the anode profile within the space bounded by the cathode are ordered in dimensionless time $\tau$. Until the GV surface arrives at the point $P_A(\tilde{R}_A(\tilde{Z}), \tilde{Z})$, the curves obeying equation (41) that begin from that point are not recognized as characteristics curves within the meaning of the Method of Characteristics [23], because until the time $\tau \geq \tau(\tilde{Z})$, they are not in the neighborhood of the solution of the equation.

The constant $C_1$ for each characteristic curve is then given by

$$C_{1,A}(\tilde{Z}, N) = \frac{\tilde{Z}}{N_A(\tilde{Z})} - s \text{ArcCosh}\left(\frac{\tilde{R}_A(\tilde{Z})}{|N_A(\tilde{Z})|}\right) \qquad (51)$$

Varying $\tilde{r}$ in (41) from $\tilde{R}_A(\tilde{Z})$ to $\tilde{r}_C$ with $N = N_A(\tilde{Z})$ generates the characteristic curve $\tilde{r}_{ch} \equiv (\tilde{r}, \tilde{z}_{ch})$ for point $P_A(\tilde{R}_A(\tilde{Z}), \tilde{Z})$. At the vertices $(\tilde{R}_A(\tilde{Z}_A^{(i)}), \tilde{Z}_A^{(i)})$, $d\tilde{R}_A/d\tilde{Z}$ may be discontinuous. Characteristics at these points are found by using

$$C_{1,A}\left(\tilde{Z}_A^{(i)}\right) = \frac{\tilde{Z}_A^{(i)}}{N} - s\text{ArcCosh}\left(\frac{\tilde{R}_A\left(\tilde{Z}_A^{(i)}\right)}{|N|}\right) \tag{52}$$

by varying N continuously from

$$N_A^-\left(\tilde{Z}_A^{(i)}\right) = \tilde{R}_A\left(\tilde{Z}_A^{(i)}\right)\Big/\sqrt{1+\left(d\tilde{R}_A^{(i-1)}(\tilde{Z})/\tilde{Z}\right)^2_{\tilde{Z}\to\tilde{Z}_A^{(i)}}} \tag{53}$$

to

$$N_A^+\left(\tilde{Z}_A^{(i)}\right) = \tilde{R}_A\left(\tilde{Z}_A^{(i)}\right)\Big/\sqrt{1+\left(d\tilde{R}_A^{(i)}(\tilde{Z})/\tilde{Z}\right)^2_{\tilde{Z}\to\tilde{Z}_A^{(i)}}}. \tag{54}$$

Each characteristic is therefore labeled *both* by its starting point $P_A\left(\tilde{R}_A(\tilde{Z}),\tilde{Z}\right)$ and its N value as $\text{Ch}(\tilde{Z},N)$. For smooth profiles, with continuous first derivatives, $N_A(\tilde{Z})$ does not have discontinuities at the vertices.

Similarly, every point $P_I\left(\tilde{R}_I(\tilde{Z}),\tilde{Z}\right)$ on the insulator generates a characteristic curve whose constant $C_1$ is given by

$$C_{1,\text{In}}\left(\tilde{Z}\right) = \pm\frac{\tilde{Z}\sqrt{\left(d\tilde{R}_I(\tilde{Z})/d\tilde{Z}\right)^2+1}}{\tilde{R}_I(\tilde{Z})\left(d\tilde{R}_I(\tilde{Z})/d\tilde{Z}\right)} - s\text{ArcCosh}\left(\frac{\sqrt{\left(d\tilde{R}_I(\tilde{Z})/d\tilde{Z}\right)^2+1}}{\left|\left(d\tilde{R}_I(\tilde{Z})/d\tilde{Z}\right)\right|}\right) \tag{55}$$

It is well known [23] that a single-valued, continuous and differentiable (aka classical) solution to a nonlinear first order partial differential equation exists if, and only if, the characteristic curves cover the entire domain and do not intersect. For arbitrary anode and insulator shapes, this circumstance is not guaranteed. Intersection of characteristic curves or their exclusion from certain regions of the solution domain leads to non-classical solutions that represent a system of shock waves and rarefaction waves [23] where additional physical conditions such as the "entropy condition" or a "vanishing viscosity" need to be invoked to decide which is a "physical solution".

However, if the intersecting curves of family $\mathbb{C}$ are associated with different values of $\tau$, then those that correspond to earlier time are to be considered the "genuine" characteristics and those that correspond to later time are "forbidden to intrude in the region" occupied by the former. *This is because the characteristic curves are collection of characteristic line elements which are defined only with respect to the time-dependent solution of the GV equation.* This process prevents generation of intersecting *characteristics* even though intersecting *curves* obeying the algebraic equation (41) may be drawn over the anode profile. The intersecting curves that satisfy equation (41) may be compared with two intersecting trajectories of solid objects which do not collide because they arrive at the point of intersection of the trajectories at different times. The solution region corresponding to each segment of external branch of anode profile should thus comprise of three (or more) sub-regions: one which is based on characteristics drawn from the points on the external branch of anode profile in that segment, second which is based on characteristics drawn from the vertex at the beginning of the segment and the third that is based on characteristics drawn from points in the previous anode segment or insulator (with additional regions based on characteristics drawn from still earlier segments). This is illustrated in Section III. The characteristics associated with the internal branch of anode profile do not intersect with those associated with the external branch except at the transition zone between the two.

For the purpose of the present discussion, it is assumed that anode and insulator profiles have been chosen in such manner that the characteristics associated with the same value of time do not intersect and that a classical solution to the GV equation exists in each sub-region of the solution space as discussed above. As an example of a situation that does not conform with this assumption, consider a square-cross-section groove cut into the anode in the rundown phase. A classical solution to the GV equation may not exist in this case: the solution is likely to jump across the groove. However, this kind of situation is outside the scope of the present discussion.

The GV surface at dimensionless time $\tau$ is the locus of those points on the entire family $\mathbb{C}$ of characteristic curves, whose radial coordinates satisfy equation (42), and which continuously connect the anode with the cathode. The dimensionless time $\tau(\tilde{Z})$ at which the intersection of the GV surface with the anode reaches the point $P_A(\tilde{R}_A(\tilde{Z}), \tilde{Z})$ has already been determined above.

The constant $C_2$ in (42) for the characteristic $Ch(\tilde{Z}, N)$ is then given by

$$C_{2,A}(\tilde{Z}, N) = \frac{\tilde{R}_A(\tilde{Z})}{|N|} \sqrt{\frac{\tilde{R}_A^{\ 2}(\tilde{Z})}{N^2} - 1} + \text{ArcCosh}\left(\frac{\tilde{R}_A(\tilde{Z})}{|N|}\right) - \frac{s\tau(\tilde{Z})}{N^2} \qquad (56)$$

At any time $\tau > \tau(\tilde{Z})$, the intersection of the GV surface with the characteristic $\text{Ch}(\tilde{Z}, N)$ can be calculated by solving equation (42) for $\tilde{r} \equiv \tilde{r}_{GV}(\tau, \tilde{Z}, N)$, with $C_2$ given by (56) and substituting in (41) for obtaining $\tilde{z} \equiv \tilde{z}_{GV}(\tau, \tilde{Z}, N)$. The set of all such points for a given $\tau$ is the GV surface.

A convenient way of visualizing the solution space of the GV equation is to consider the set of all points calculated by varying values of $N(\tilde{Z})$ and $\tau$. If these points are grouped by varying $N(\tilde{Z})$ and keeping $\tau$ constant, each group will be a GV surface at that constant value of $\tau$. If these points are grouped by varying $\tau$ and keeping a constant value of $N(\tilde{Z})$, each group will be a characteristic corresponding to that constant value of $N(\tilde{Z})$.

GV have given the following method of calculating the GV surface. Rewrite (41), the equation for the characteristic $\text{Ch}(\tilde{Z}, N)$, defining $\alpha/2 = C_1 - \tilde{z}/N$, $\beta/2 \equiv \text{ArcCosh}(\tilde{R}_A(\tilde{Z})/N)$ as

$$\begin{aligned} \tilde{r}_{GV}(\alpha, N) &= |N|\text{Cosh}(\alpha/2); \\ \tilde{z}_{GV}(\alpha, N, s) &= NC_1(\tilde{Z}, N) + Ns\alpha/2 \end{aligned} \qquad (57)$$

Then (42) can be written in three different ways as

$$F(\alpha) \equiv \text{Sinh}(\alpha) + \alpha = 2\left(C_2(\tilde{Z}, N, s) + \frac{s\tau}{N^2}\right) \qquad \cdots(a)$$

$$= 2\left(\frac{\tilde{R}_A(\tilde{Z})}{|N(\tilde{Z})|} \sqrt{\frac{\tilde{R}_A(\tilde{Z})^2}{N^2(\tilde{Z})} - 1} + \text{ArcCosh}\left(\frac{\tilde{R}_A(\tilde{Z})}{|N(\tilde{Z})|}\right) + s\frac{(\tau - \tau(\tilde{Z}))}{N^2(\tilde{Z})}\right) \qquad \cdots(b) \quad (58)$$

$$= \text{Sinh}(\beta) + \beta + \text{Cosh}^2(\beta/2)\frac{2s(\tau - \tau(\tilde{Z}))}{\tilde{R}_A(\tilde{Z})^2} \qquad \cdots(c)$$

Note that $F(\alpha)$ is invariant under the transformation $N \to -N$ and the function $F(\alpha)$ is an odd function of its argument. GV provide [1] the following inversion formula for $|F(\alpha)|$

$$|\alpha| = \frac{|F|}{2}\left(1 - \frac{|F|^2}{48}\right) \quad \text{for} \quad |F| < 1$$

$$|\alpha| = \text{Log}\left[2|F| + \frac{1}{2|F|} - 2\left(1 - \frac{1}{|F|}\right)\text{Log}(2|F|)\right] \quad \text{for} \quad |F| \gg 1 \tag{59}$$

GV quote an error of less than 0.1% in the second formula for $|F| > 5$. For the entire set of characteristics $\text{Ch}(\tilde{Z}, N)$, $\alpha$ can be determined as a function of $\tau$ from equation (58)(a) from which the GV surface $(\tilde{r}_{GV}, \tilde{z}_{GV})$ is computed from (57).

For the special case of constant anode radius (e.g. rundown phase in the Mather type plasma focus), (47) gives $N_A(\tilde{Z}) = \tilde{R}_A$. Then $C_{1,A}(\tilde{Z}, N) = \tilde{Z}/\tilde{R}_A$, $C_{2,A}(\tilde{Z}, N) = -s\tau(\tilde{Z})/\tilde{R}_A^2$. From (49) , $\tau(\tilde{Z}) = 2\tilde{R}_A(\tilde{Z} - \tilde{Z}_A^{(i)}) + \tau_i$. This can be used to eliminate $\tilde{Z}$ in (41) and (42) giving

$$\tilde{z} = \tilde{Z}_A^{(i)} + \frac{1}{2\tilde{R}_A}(\tau - \tau_i) + \frac{1}{2}\tilde{R}_A \text{ArcCosh}\left(\frac{\tilde{r}}{\tilde{R}_A}\right) - \frac{1}{2}\frac{\tilde{r}}{\tilde{R}_A}\sqrt{\tilde{r}^2 - \tilde{R}_A^2} \tag{60}$$

Varying $\tilde{r}$ from $\tilde{R}_A$ to $\tilde{r}_C$ generates the GV surface in the segment with constant anode radius *if characteristics from the previous anode profile segment do not intrude into the rundown space.* Section III provides a counterexample where they do. For $\tilde{R}_A = 1$, (60) describes the rundown phase of the Mather type plasma focus [1,5,6]

The solution of the generalized plasma focus problem (i.e., GV equation along with the supplementary conditions) requires appreciation of the existence of an inherent bifurcation phenomenon. The solution consisting of locus of the coordinates $(\tilde{r}_{GV}(\alpha, N), \tilde{z}_{GV}(\alpha, N, s))$ defined by (57) and (58) becomes double-valued (i.e. it breaks up into two branches) beyond a given value of $\tau$. This can arise in two kinds of cases

From (40) and (57) it is clear that $d\tilde{r}/d\tilde{z} = sN^{-1}\sqrt{\tilde{r}^2 - N^2} = 0$ when $\alpha = 0$ showing that $\tilde{r}$ has a minimum with respect to $\tilde{z}$. Due to a monotonic variation given by (58), $\alpha$ may cross zero and change its sign. Beyond the condition defined by $\alpha = 0$ the solution becomes double valued: which correspond to $\tilde{r} = N\text{Cosh}(|\alpha|/2)$, $\tilde{z} = NC_1(\tilde{Z},N) \pm Ns|\alpha|/2$. From (58)

$$\alpha = 0 \Rightarrow \frac{\tilde{R}_A(\tilde{Z})}{|N(\tilde{Z})|}\sqrt{\frac{\tilde{R}_A(\tilde{Z})^2}{N^2(\tilde{Z})} - 1} + \text{ArcCosh}\left(\frac{\tilde{R}_A(\tilde{Z})}{|N(\tilde{Z})|}\right) = -s\frac{(\tau - \tau(\tilde{Z}))}{N^2(\tilde{Z})} \quad (61)$$

If a real solution $N \equiv N_{br}(\tau - \tau(\tilde{Z}))$ to (61) exists, the coordinates $(\tilde{r}_{br}, \tilde{z}_{br})$ of the branch points are given by

$$\tilde{r}_{br} = N_{br}(\tau - \tau_i); \quad \tilde{z}_{br} = \tilde{Z}_A^{(i)} + N_{br}(\tau - \tau_i)\text{ArcCosh}\left(\frac{\tilde{R}_A(\tilde{Z}_A^{(i)})}{N_{br}(\tau - \tau_i)}\right) \quad (62)$$

The branch point radius $\tilde{r}_{br}$ is the minimum possible radius of the GV surface at $\tau$ for the concerned vertex or segment.

Another phenomenon that needs to be appreciated is that the family $\mathbb{V}$ of characteristics emitted from a vertex $(\tilde{r}_v, \tilde{z}_v)$ given by $\tilde{r} = \tilde{r}(\alpha, N); \tilde{z} = \tilde{z}(\alpha, N, s, \tilde{r}_v, \tilde{z}_v)$ has an envelope $\Sigma_v$ (a curve that is tangent to each member of the family $\mathbb{V}$) which satisfies the equation [24]

$$\partial_\alpha \tilde{r}(\alpha, N)\partial_N \tilde{z}(\alpha, N, s, \tilde{r}_v, \tilde{z}_v) - \partial_N \tilde{r}(\alpha, N)\partial_\alpha \tilde{z}(\alpha, N, s, \tilde{r}_v, \tilde{z}_v) = 0 \quad (63)$$

Equation (63) gives $\alpha_{sol}(N/\tilde{r}_v)$ as a function of $(N/\tilde{r}_v) = \text{Sech}(\beta/2)$ as a non-trivial solution of the following equation [25]

$$\text{Coth}(\alpha/2) - \frac{1}{2}\alpha = \text{Coth}(\beta/2) - \frac{1}{2}\beta \quad (64)$$

The envelope is then given by the parametric curve

$$\Sigma_v = \left(\tilde{r}(\alpha_{sol}(N/\tilde{r}_v), N), \tilde{z}(\alpha_{sol}(N/\tilde{r}_v), N, s, \tilde{r}_v, \tilde{z}_v)\right) \quad (65)$$

which explicitly demonstrates that the envelope satisfies equation (41) that defines characteristics of the GV equation and is therefore itself a characteristic curve. The envelope and the characteristics are shown in Fig 1.

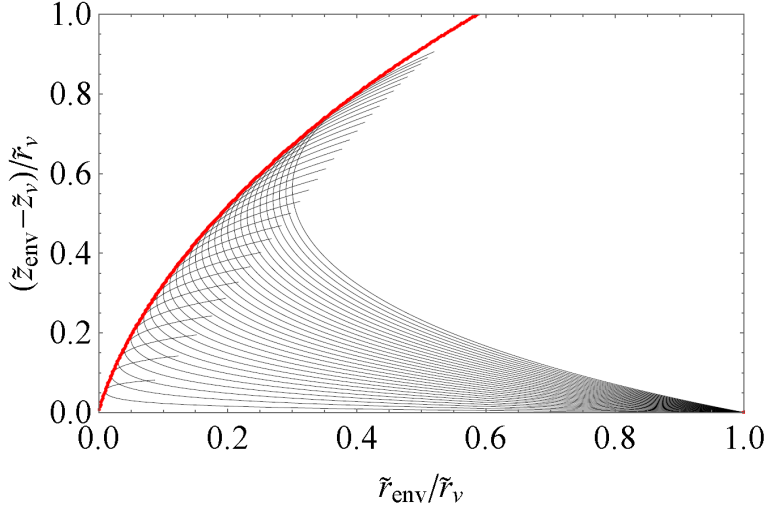

Fig. 1: Family $\mathbb{V}$ of characteristics emitted from vertex $(\tilde{r}_v, \tilde{z}_v)$ with s=-1 shown in black, parametrized by N. For each curve, N is constant while $\alpha$ varies along the curve. The envelope $\Sigma_v$ is shown as a thick red line. Note that the envelope represents the boundary of the zone of influence of the family $\mathbb{V}$ of characteristics starting from vertex $(\tilde{r}_v, \tilde{z}_v)$.

The envelope is also the locus of smallest-radius points on the family of characteristics emitted from a vertex. However, it differs from the branching phenomenon, because it is not defined by the condition $\alpha = 0$. It is also defined for a vertex, not a segment.

It is clear that the vertex $(\tilde{r}_v, \tilde{z}_v)$ influences $\Sigma_v$ in only two ways: (1) by providing a scaling of radial and axial coordinates of the envelope with $\tilde{r}_v$ and (2) by redefining origin for the axial coordinate of the envelope with $\tilde{z}_v$. The shape of the function $(\tilde{z}_{env} - \tilde{z}_v)/\tilde{r}_v$ versus $\tilde{r}_{env}/\tilde{r}_v$ is in fact universal. Fig 2 compares a power-law fit with this shape

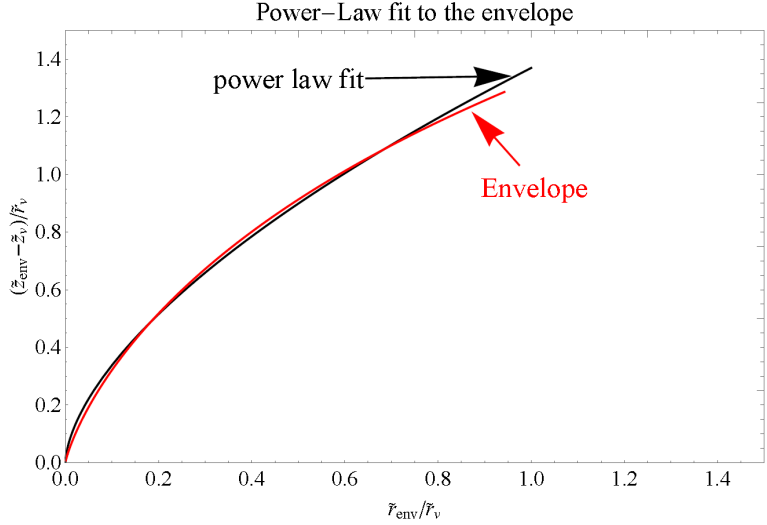

Fig 2: Power law fit to the envelope $\Sigma_v$: $(\tilde{z}_{env} - \tilde{z}_v)/\tilde{r}_v = a_1 \times (\tilde{r}_{env}/\tilde{r}_v)^{a_2}$  $a_1 = 1.36996; a_2 = 0.607746$. The fit constants change if only a part of the envelope profile is chosen to obtain a better quality of fit. For a general plasma focus problem, it is better to determine these coefficients ab initio.

In the case of the generalized plasma focus problem, the characteristics in the segment $\tilde{R}_{A,ext}^{(i)}(\tilde{Z})$, $\tilde{Z}_A^{(i-1)} \leq \tilde{Z} \leq \tilde{Z}_A^{(i)}$ are given by

$$\tilde{r}(\alpha,\tilde{Z}) = N_A(\tilde{Z})\text{Cosh}(\alpha/2)$$
$$\tilde{z}(\alpha,\tilde{Z},s) = \tilde{Z} - sN_A(\tilde{Z})\text{ArcCosh}\left(\frac{\tilde{R}_A(\tilde{Z})}{N_A(\tilde{Z})}\right) + N_A(\tilde{Z})s\alpha/2 \qquad (66)$$

where $N_A(\tilde{Z})$ is constant along the characteristic and the parameter $\alpha$ varies from its value

$$\alpha_A = 2\text{ArcCosh}\sqrt{1+(d\tilde{R}_A/d\tilde{Z})^2} \qquad (67)$$

at the anode to

$$\alpha_C = 2\text{ArcCosh}(\tilde{r}_C/N_A(\tilde{Z})) \qquad (68)$$

Unlike the case of characteristics from a vertex considered above, the two parameters of this family are $\alpha$ and $\tilde{Z}$ rather than $\alpha$ and $N$. The envelope of characteristics is then given by [23]

$$\partial_\alpha \tilde{r}(\alpha,\tilde{Z})\partial_{\tilde{Z}}\tilde{z}(\alpha,\tilde{Z},s) - \partial_{\tilde{Z}}\tilde{r}(\alpha,\tilde{Z})\partial_\alpha \tilde{z}(\alpha,\tilde{Z},s) = 0 \qquad (69)$$

whose numerical solution for a given anode profile gives $\alpha_{sol}(\tilde{Z})$ from which the envelope is obtained as

$$\Sigma_s = \left(\tilde{r}\left(\alpha_{sol}(\tilde{Z}),\tilde{Z}\right), \tilde{z}\left(\alpha_{sol}(\tilde{Z}),\tilde{Z},s\right)\right) \tag{70}$$

Since the characteristics are tangent to the anode profile, in many cases, the envelope simply turns out to be the anode profile. However, it is not known whether any characteristics other than the anode profile may exist for certain profile functions.

Since equation (37) is invariant under the transformation $N \to -N$ and (41) is invariant under the combined operation $N \to -N, \tilde{z} \to -\tilde{z}$, the mirror image $\overline{\mathbb{V}}$ of the family $\mathbb{V}$ of characteristics (shown in Fig 1) about the horizontal line through the vertex $(\tilde{r}_v, \tilde{z}_v)$ are also characteristics of the GV equation.

The properties of the GV surface in the limit $N \to 0$ merit a special discussion. One important example of such case is the neighborhood of the axis. From (57), it is clear that the GV surface can reach the axis at $\tilde{r} = 0$ only when $N \to 0$, since the hyperbolic cosine function cannot become zero for real arguments. However, the right-hand side of (58)(b) diverges as $N \to 0$. This fact causes numerical instabilities that need to be handled appropriately or avoided by not approaching the neighborhood of N=0 too closely. It has the following series expansion in powers of N

$$\frac{2\left(\left(\tilde{R}_A(\tilde{Z})\right)^2 + s\left(\tau - \tau(\tilde{Z})\right)\right)}{N^2} - 1 + \log\left[\frac{4\left(\tilde{R}_A(\tilde{Z})\right)^2}{N^2}\right] - \frac{3N^2}{4\left(\tilde{R}_A(\tilde{Z})\right)^2} - \frac{5N^4}{16\left(\tilde{R}_A(\tilde{Z})\right)^4} + O[N]^6 \tag{71}$$

showing that there is an essential singularity that vanishes when $s = -1$ (radial velocity directed towards the axis) and

$$\tau \to \tau(\tilde{Z}) + \left(\tilde{R}_A(\tilde{Z})\right)^2 \equiv \tau_p \tag{72}$$

Additionally, there is a logarithmic singularity. Since the right hand side of (58) diverges as $N \to 0$, the second expression of (59) applies and can be approximated for large $|F(\alpha)|$ as $|\alpha| \approx \text{Log}[|2F|]$, $\text{Cosh}(\alpha/2) \approx \frac{1}{2}\exp(\alpha/2) \approx \sqrt{|F|/2}$. Then (57) gives

$$\tilde{r}^2 = N^2 \text{Cosh}^2(\alpha/2)$$

$$\approx \left| \begin{array}{l} \left(\tilde{R}_A(\tilde{Z})\right)^2 + s\left(\tau - \tau(\tilde{Z})\right) \\ -N^2 + N^2 \log\left[\dfrac{4\left(\tilde{R}_A(\tilde{Z})\right)^2}{N^2}\right] - N^2 \left\{ \dfrac{3N^2}{4\left(\tilde{R}_A(\tilde{Z})\right)^2} + \dfrac{54N^4}{16\left(\tilde{R}_A(\tilde{Z})\right)^4} + O[N]^7 \right\} \end{array} \right| \quad (73)$$

Two cases need to be discussed separately: (I) Type I anode profile where the normalized radius of the last "flat-top" segment of the external branch of anode profile (before beginning of the cavity if any), $\tilde{r}_{\text{last}} = \tilde{R}_A^{(\text{last})}\left(\tilde{Z}_A^{(\text{last})}\right)$ is assumed to be non-zero (it may be arbitrarily small). Note that according to the discussion of (47), N has negative values in the cavity. (II) Type II anode profile where $R_A^{(\text{last})}(\tilde{Z}) \to 0$ as $\tilde{Z} \to \tilde{z}_A$. The difference in the two cases lies in the fact that $\tilde{Z}$ ceases to be a good variable to describe points on the flat top anode in equations (51) or (56) in the first case but not the second case. The characteristics that enable the GV surface to reach the axis are emitted from the last vertex in the first case but are emitted from each point on the anode profile right up to the axis. So for the first case, functions $\tilde{R}_A(\tilde{Z})$ and $\tau(\tilde{Z})$ are replaced with constants $\tilde{r}_{\text{last}}$ and $\tau_{\text{last}}$ in (73). From (47), it is seen that the ratio $\tilde{R}_A(\tilde{Z})/N$ in (73) has a finite limit in the second case as $N \to 0$.

In both cases, it is clear that the last 3 terms of (73) vanish in the limit of $N \to 0$ but the variation of $\tilde{r}^2$ with sufficiently small values of N is different for the two cases. The result for $N \to 0$ is

$$\tilde{r}^2 = \left|\tilde{r}_{\text{last}}^2 + s(\tau - \tau_{\text{last}})\right| \quad \text{or} \quad \tilde{r}^2 = \left|\left(\tilde{R}_A(\tilde{Z})\right)^2 + s(\tau - \tau(\tilde{Z}))\right|, \quad \tilde{Z}_A^{(i_m-1)} \leq \tilde{Z} \leq \tilde{Z}_A^{(i_m)} \quad (74)$$

Since $\tau > \tau_{last}$ (or $\tau(\tilde{Z}_A^{(i_m-1)})$), the axis is approached only when $s = -1$ (when the radial velocity is directed towards the axis). Also, from (57) and (52), $\tilde{z} \to \tilde{z}_A$ as $N \to 0$. This shows that *in the plane $\tilde{z} = \tilde{z}_A$ containing the top of the anode, the GV surface resembles contracting or expanding circles centered at the axis*. This also follows from equation (43). (Note that the divergent behavior in the limit $N \to 0$ causes numerical inaccuracies so that the radius does not approach zero at the exact value of $\tau_p$ mentioned above. In fact, the neighborhood of the axis is very difficult to approach numerically. So extreme caution must be exercised when using numerical results that involve N=0)

The fact that identical non-zero values of the radial coordinate occur at two values of $\tau$ implies that the GV surface gets reflected from the axis at $\tau = \tau_{last} + \tilde{r}_{last}^2 \equiv \tau_p$ for the first case or $\tau = \tau(\tilde{z}_A) \equiv \tau_p$ for the second case. *This follows purely from the mathematical properties of the GV equation and has nothing to do with the physics of shock wave reflection from a symmetry axis that involves conservation laws for mass, momentum and energy and appropriate boundary conditions at the axis.*

This reflection is in fact an instance of the bifurcation phenomenon described earlier, with the branching point at the axis in the plane containing the vertex $\tilde{Z}_A^{(last)}$. All subsequent branch points lie on the envelope of characteristics emitted from $\tilde{Z}_A^{(last)}$ for a Type I anode profile. The corresponding time $\tau_{br}$ can be obtained from (58) using the solution $\alpha_{sol}(N/\tilde{r}_{last})$ of (64):

$$\left(\tau_{br} - \tau_{last} - \tilde{r}_{last}^2\right)$$
$$= \tilde{r}_{last}^2 \left\{ \sqrt{1 - \frac{N^2}{\tilde{r}_{last}^2}} - 1 + \frac{N^2}{\tilde{r}_{last}^2} \text{ArcCosh}\left(\frac{\tilde{r}_{last}}{N}\right) - \frac{N^2}{\tilde{r}_{last}^2} \frac{1}{2}\left(\text{Sinh}(\alpha_{sol}(N/\tilde{r}_{last})) + \alpha_{sol}(N/\tilde{r}_{last})\right) \right\} \quad (75)$$

The function in braces has a good regression fit to a 3rd degree polynomial in $(N/\tilde{r}_{last})$ over the range $[0, 0.5]$ with coefficients {0, 0.454231, -0.879163, 8.11138}.

The lower branch spanning $0 \le N \le N_0(\tau_{br} - \tau_p)$, where $N = N_0(\tau_{br} - \tau_p)$ is the solution of (75), appears as a Radially Expanding Front (REF), that has the radius given by (74) in the plane containing the apex of the anode, which must additionally meet the anode. The upper branch spanning $N_0(\tau - \tau_p) \leq N \leq \tilde{r}_{last}$ forms part of the GV surface that connects the REF to the cathode.

For a Type II anode profile, branching of the GV surface occurs for $\tau > \tau_p$ but the locus of branch points is not given by (75). Since the details of the anode profile in the last segment play a role in determining the locus of branch points, one can simply determine it empirically for each case rather than analytically for a general profile, as is done in the illustration in Section III.

The space between the locus of branch points and the axis is not reached by any characteristic either from the vertex $\tilde{Z}_A^{(last)}$ for Type I anode profile or the last segment of the Type II anode profile. Since the envelope from the last vertex is shown to be a characteristic within the meaning of (41), which is tangent to the axis and additively dependent on the axial coordinate of the vertex, it can be used as a template for constructing a family $\mathbb{S}$ of characteristics by translating it along the axis with the free parameter $\tilde{z}_0$:

$$\tilde{\tilde{z}}_S = \tilde{\tilde{z}}_0 \pm a_1 \tilde{\tilde{r}}_S^{a_2} \tag{76}$$

The double overtilde in (76) denotes coordinates normalized to $\tilde{r}_{last}$. The negative sign of $a_1$ applies to the mirror image $\overline{\mathbb{S}}$ which would also be a family of characteristics in view of the invariance properties of (41) mentioned earlier. The mirror image $\overline{\mathbb{S}}$ would be used when considering PCS movement into a cavity in the anode.

The GV surface in the space between the locus of branch points and the axis then consists of a family $\mathbb{O}$ (or its mirror image $\overline{\mathbb{O}}$) of orthogonal curves of the family $\mathbb{S}$ (or its mirror image $\overline{\mathbb{S}}$) given by

$$\tilde{\tilde{z}}_{GV} = a_3 \mp \frac{1}{a_1 a_2 (2 - a_2)} \tilde{\tilde{r}}_{GV}^{2-a_2} \tag{77}$$

where the positive sign applies for PCS movement into a cavity in the anode. The constant $a_3$ is determined by matching (77) at the branch point given by (62) or otherwise determined for a given anode profile. The family $\mathbb{O}$ describes an Axially Expanding Front (AEF). The significance of the mirror family $\overline{\mathbb{O}}$ is discussed in Section IV.

For times $\tau > \tau_p$, the GV surface therefore consists of *three* branches that intersect the locus of branch points at $(\tilde{r}_{br}, \tilde{z}_{br})$: a Radially Expanding Front (REF), whose intersection with the anode has radius given by (74), which is orthogonal to the family $\mathbb{C}$ of characteristics; the Axially Expanding Front (AEF) that is orthogonal to the family $\mathbb{S}$ and a third branch that is orthogonal to $\mathbb{C}$ but which connects the point of intersection to the cathode (referred as CN for connector), which continues to move along its normal. This situation is illustrated in Fig 3 of Ref. 5 for a Mather type plasma focus and also illustrated in Section III.

For a Type II anode profile, the REF would be moving towards lower values of $\tilde{Z}$ so that $S = -1$ in (47) and (49). One can define a "ghost segment" on the external branch of anode profile for computational purposes to deal with the GV surface for $\tau > \tau_p$ with the sign of $N_A(\tilde{Z})$ reversed. It should be noted that physical considerations of supplementary condition 3 do not apply to the reflected GV surface and there is no necessity that the reflected GV surface should intersect with the anode at right angles.

The AEF and REF define an enclosed volume that expands with dimensionless time $\tau > \tau_p$. Fig 3 illustrates this for the Mather geometry. Incidentally, $(\tilde{r}_{br}, \tilde{z}_{br})$ also represent the radius and height of the largest cylinder that can be inscribed in this closed volume. The point of intersection of the AEF and REF, $(\tilde{r}_{br}, \tilde{z}_{br})$, moves along the line shown in blue according to the parametric relation

$$\tilde{r}_{br} = 0.8143(\tau - \tau_p)^{0.59451}; \tilde{z}_{br} = 1.320(\tau - \tau_p)^{0.39898} \qquad (78)$$

Along the axis, the position of the AEF is given by the regression fit

$$\tilde{z}_{APF} = 1.70246(\tau - \tau_p)^{0.443703} \qquad (79)$$

This result will be used in Section IV.

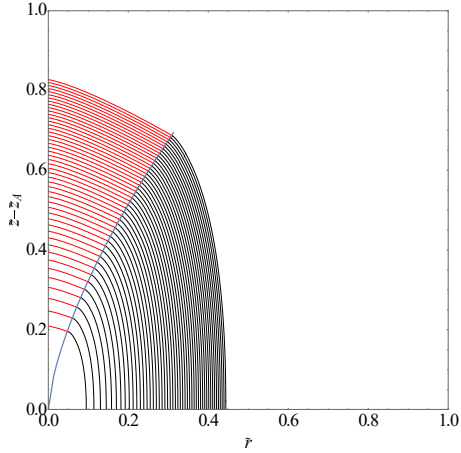

Fig 3.: The Axially Expanding Front (AEF), shown in red, and the Radially Expanding Front (REF), shown in black, define a closed volume that steadily expands for $\tau > \tau_p$. The figure is for the Mather geometry case. For any other case, the coordinates are scaled to the radius of the last vertex.

The points on the AEF in Fig 3, are well-represented by the following regression fits

$$\tilde{r}_{APF} = (0.82227 - 0.00814\,j)(\tau - \tau_p)^{0.59392}$$
$$\tilde{z}_{APF} = (1.31164 + 0.00507\,j - 0.00001117\,j^2)(\tau - \tau_p)^{0.3982 + 0.0006571\,j - 0.000001994\,j^2} \quad (80)$$

with j varying from 1 on the blue line separating the AEF and REF to 101 on the axis. No simple regression fit is available for the REF at this time.

The interesting point about this phenomenon is that the expanding enclosed volume has *no physics* associated with it other than what is built into the GV equation. It could be used to model the expansion of the DPF plasma after the pinch phase without invoking pressure balance and energy balance considerations, just to verify whether the experimentally observed plasma expansion is really sensitive to some physical assumptions or not.

The practical implementation and utility of this generalization of the GV model is illustrated with concrete examples in Part III of this series. It can be readily verified that the case of the classical Mather type plasma focus discussed earlier [1,5,6] is a special case of the above method, with 'a' chosen as the anode radius.

The above discussion of the properties of the GV surface applies for all values of the generalization constant $\chi$, which is absorbed in the definition of $\tau$, but can be written explicitly when needed, as discussed later. This allows a family of GV surfaces to be constructed with $\chi$ as an additional parameter, which shall be referred as $\chi$–surfaces. The additional degree of freedom allows the $\chi$–surfaces to have local normal velocity $\chi$ times the local normal velocity of the GV surface.

### III. Tutorial illustrations of the kinematic formalism

This section offers a tutorial illustration of the procedure outlined above using both type I and type II profiles, the latter one taken first because of its significant departure from a conventional profile. Mathematica notebook files for these examples are available from the author on reasonable request.

#### a. Illustration with a non-standard Type-II geometry

The type II profile of the form described below is assumed for its following features: (1) multi-segment character for both insulator and anode profiles (2) discontinuous nature of $N(\tilde{Z})$ (3) its simple, practically implementable form.

$$\begin{aligned}
\tilde{R}_I(\tilde{Z}) &= \tilde{R}_I^{(0)} + \left(\tilde{R}_I^{(1)} - \tilde{R}_I^{(0)}\right)\tilde{Z}/\tilde{Z}_I^{(1)} & 0 \leq \tilde{Z} \leq \tilde{Z}_I^{(1)} &\Leftrightarrow I_1 \\
&= \tilde{R}_I^{(1)} + \left(\tilde{Z} - \tilde{Z}_I^{(1)}\right)\left(\tilde{R}_I^{(2)} - \tilde{R}_I^{(1)}\right)/\left(\tilde{Z}_I^{(2)} - \tilde{Z}_I^{(1)}\right) & \tilde{Z}_I^{(1)} \leq \tilde{Z} \leq \tilde{Z}_I^{(2)} &\Leftrightarrow I_2
\end{aligned} \tag{81}$$

$$\begin{aligned}
\tilde{R}_A(\tilde{Z}) &= \tilde{R}_A^{(1)} = \tilde{R}_I^{(2)} & \tilde{Z}_A^{(0)} \leq \tilde{Z} \leq \tilde{Z}_A^{(1)}; \tilde{Z}_A^{(0)} = \tilde{Z}_I^{(2)} = \tilde{z}_I &\Leftrightarrow A_1 \\
&= \tilde{R}_A^{(1)} + \left(\tilde{Z} - \tilde{Z}_A^{(1)}\right)\left(\tilde{R}_A^{(2)} - \tilde{R}_A^{(1)}\right)/\left(\tilde{Z}_A^{(2)} - \tilde{Z}_A^{(1)}\right) & \tilde{Z}_A^{(1)} \leq \tilde{Z} \leq \tilde{Z}_A^{(2)} &\Leftrightarrow A_2 \\
&= \tilde{R}_A^{(2)} & \tilde{Z}_A^{(2)} \leq \tilde{Z} \leq \tilde{Z}_A^{(3)} &\Leftrightarrow A_3 \\
&= \tilde{R}_A^{(2)}\left(\tilde{Z}_A^{(4)} - \tilde{Z}\right)/\left(\tilde{Z}_A^{(4)} - \tilde{Z}_A^{(3)}\right) & \tilde{Z}_A^{(3)} \leq \tilde{Z} \leq \tilde{Z}_A^{(4)} &\Leftrightarrow A_4
\end{aligned} \tag{82}$$

The profile is illustrated in Fig 4 along with the GV surfaces up to $\tau_p$.

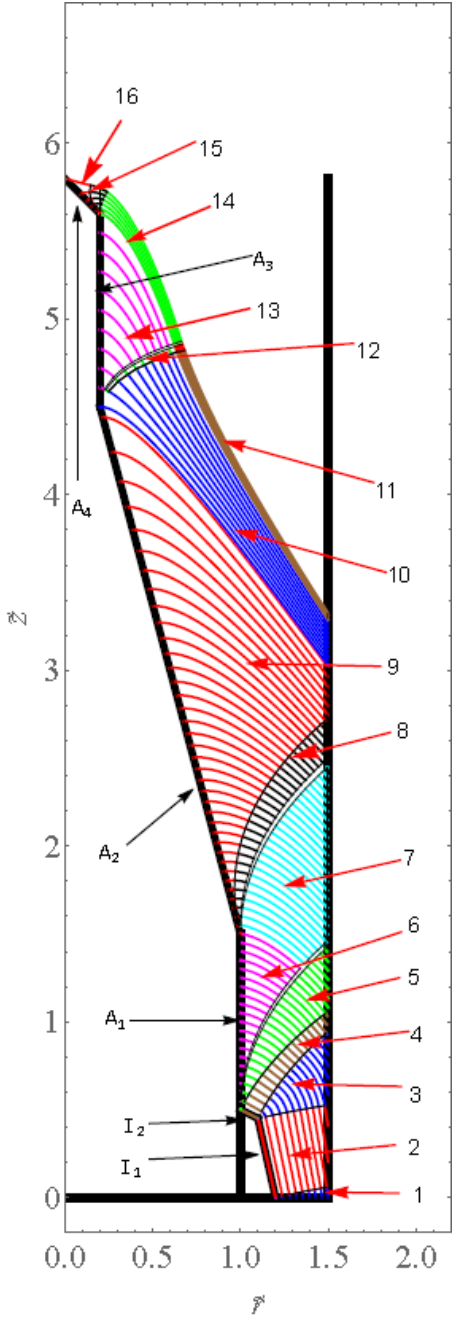

Fig 4: GV surfaces as a function of $\tau \leq \tau_p$. Device profile described by (82) shown in thick black straight lines. The parameters are as follows

$\tilde{R}_I^{(0)} = 1.2$, $\tilde{R}_I^{(1)} = 1.1$, $\tilde{Z}_I^{(1)} = 0.45$, $\tilde{Z}_A^{(0)} = \tilde{Z}_I^{(2)} = \tilde{z}_I = 0.5$
$\tilde{R}_A^{(1)} = \tilde{R}_I^{(2)} = 1$; $\tilde{Z}_A^{(1)} = 1.5$, $\tilde{R}_A^{(2)} = 0.2$, $\tilde{Z}_A^{(2)} = 4.5$, $\tilde{Z}_A^{(3)} = 5.6$,
$\tilde{Z}_A^{(4)} = 5.8$ $\tilde{r}_C = 1.5$. This gives $\tau_1 = 2.0$, $\tau_2 = 5.7258$, $\tau_3 = 6.1658$, $\tau_4 = 6.22237$. The GV surfaces are shown as functions of $\tau$, in intervals of $\Delta\tau = 0.1$ up to $\tau_2$, in intervals of $\Delta\tau = (\tau_3 - \tau_2)/10 = 0.044$ in anode segment A3 and $\Delta\tau = (\tau_4 - \tau_3)/5 = 0.011314$ in anode segment A4. They are composed of 16 sub-regions, labeled with numerals and shown with different colors. Table I summarizes the parameters that are used to calculate them. The $(\tilde{r}, \tilde{z})$ values are matched at the boundaries of the sub-regions and at the cathode.

If the vertex $\tilde{Z}_I^{(1)}$ on the insulator profile is moved down to lie on the line joining the vertex at $\tilde{z} = 0$ to that at $\tilde{Z} = \tilde{z}_I$, the two segments of insulator profile merge so that sub-region 3 of Fig 4 vanishes. If it is moved further down, the two segments again become distinct but their characteristics intersect at $\tau = 0$ violating the uniqueness condition. In that case, the non-existence of a classical solution may manifest itself in a discharge that proceeds from the foot of the first insulator segment to the top of the second insulator segment bypassing the intermediate vertex $\tilde{Z}_I^{(1)}$. Contrast this with the fact that characteristics drawn from different segments of anode profile do intersect but since they become "genuine" characteristics at different times, this merely implies that the characteristics from earlier segments influence the GV surface at a time much later than the time at which the intersection of GV surface with the anode crosses the segment boundary.

Table I: Parameters for calculating GV surfaces in Fig 4.

| # | Segment /Vertex | $\tilde{Z}$ | N | $\tau$ | s |
|---|---|---|---|---|---|
| 1 | $\left(\tilde{R}_I^{(0)},0\right)$ | 0 | $\left(0\cdots N_I\left(\tilde{Z}_I^{(1)}\right)\right)$ | $(0\cdots\tau_1)$ | 1 |
| 2 | $I_1$ | $\left(0\cdots\tilde{Z}_I^{(1)}\right)$ | $N_I(\tilde{Z})$ | $(0\cdots\tau_1)$ | 1 |
| 3 | $\left(\tilde{R}_I^{(1)},\tilde{Z}_I^{(1)}\right)$ | $\tilde{Z}_I^{(1)}$ | $\left(N_I^-\left(\tilde{Z}_I^{(1)}\right)\cdots N_I^+\left(\tilde{Z}_I^{(1)}\right)\right)$ | $(0\cdots\tau_1)$ | 1 |
| 4 | $I_2$ | $\left(\tilde{Z}_I^{(1)}\cdots\tilde{z}_I\right)$ | $N_I(\tilde{Z})$ | $(0\cdots\tau_1)$ | 1 |
| 5 | $\left(\tilde{R}_A^{(1)},\tilde{z}_I\right)$ | $\tilde{z}_I$ | $\left(N_I\left(\tilde{Z}_I^{(2)}\right)\cdots N_A\left(\tilde{Z}_I^{(2)}\right)\right)$ | $(0\cdots\tau_2)$ | 1 |
| 6 | $A_1$ | $\left(\tilde{z}_I\cdots\tilde{Z}_A^{(1)}\right)$ | $N_A(\tilde{Z})$ | $(0\cdots\tau_1)$ | 1 |
| 7 | $A_1$ extended | $\left(\tilde{z}_I\cdots\tilde{Z}(\tau)\right)$ | $N_A(\tilde{Z})$ | $(\tau_1\cdots\tau_2)$ | 1 |
| 8 | $\left(\tilde{R}_A^{(1)},\tilde{Z}_A^{(1)}\right)$ | $\tilde{Z}_A^{(1)}$ | $\left(N_A^-\left(\tilde{Z}_A^{(1)}\right)\cdots N_A^+\left(\tilde{Z}_A^{(1)}\right)\right)$ | $(\tau_1\cdots\tau_2)$ | -1 |
| 9 | $A_2$ | $\left(\tilde{Z}_A^{(1)}\cdots\tilde{Z}(\tau)\right)$ | $N_A(\tilde{Z})$ | $(\tau_1\cdots\tau_2)$ | -1 |
| 10 | $A_2$ extended | $\left(\tilde{Z}_A^{(1)}\cdots\tilde{Z}(\tau)\right)$ | $N_A(\tilde{Z})$ | $(\tau_2\cdots\tau_3)$ | -1 |
| 11 | $A_2$ extended | $\left(\tilde{Z}_A^{(1)}\cdots\tilde{Z}(\tau)\right)$ | $N_A(\tilde{Z})$ | $(\tau_3\cdots\tau_4)$ | 1 |
| 12 | $\left(\tilde{R}_A^{(2)},\tilde{Z}_A^{(2)}\right)$ | $\tilde{Z}_A^{(2)}$ | $\left(N_A^-\left(\tilde{Z}_A^{(2)}\right)\cdots N_A^+\left(\tilde{Z}_A^{(2)}\right)\right)$ | $(\tau_2\cdots\tau_4)$ | -1 |
| 13 | $A_3$ | $\left(\tilde{Z}_A^{(2)}\cdots\tilde{Z}(\tau)\right)$ | $N_A(\tilde{Z})$ | $(\tau_2\cdots\tau_3)$ | -1 |
| 14 | $A_3$ extended | $\left(\tilde{Z}_A^{(2)}\cdots\tilde{Z}(\tau)\right)$ | $N_A(\tilde{Z})$ | $(\tau_3\cdots\tau_4)$ | 1 |
| 15 | $\left(\tilde{R}_A^{(2)},\tilde{Z}_A^{(3)}\right)$ | $\tilde{Z}_A^{(3)}$ | $\left(N_A^-\left(\tilde{Z}_A^{(3)}\right)\cdots N_A^+\left(\tilde{Z}_A^{(3)}\right)\right)$ | $(\tau_3\cdots\tau_4)$ | 1 |
| 16 | $A_4$ | $\left(\tilde{Z}_A^{(3)}\cdots\tilde{Z}(\tau)\right)$ | $N_A(\tilde{Z})$ | $(\tau_3\cdots\tau_4)$ | 1 |

Fig 5 demonstrates the reflection of the GV surface for a Type II anode profile. The figure shows sub-regions 14, 15 and 16 described in Table I and Fig 4 in time steps of $\Delta\tau = (\tau_4 - \tau_3)/40 = 0.00141421$ from $\tau_3$ to $\tau_4 + 0.02$.

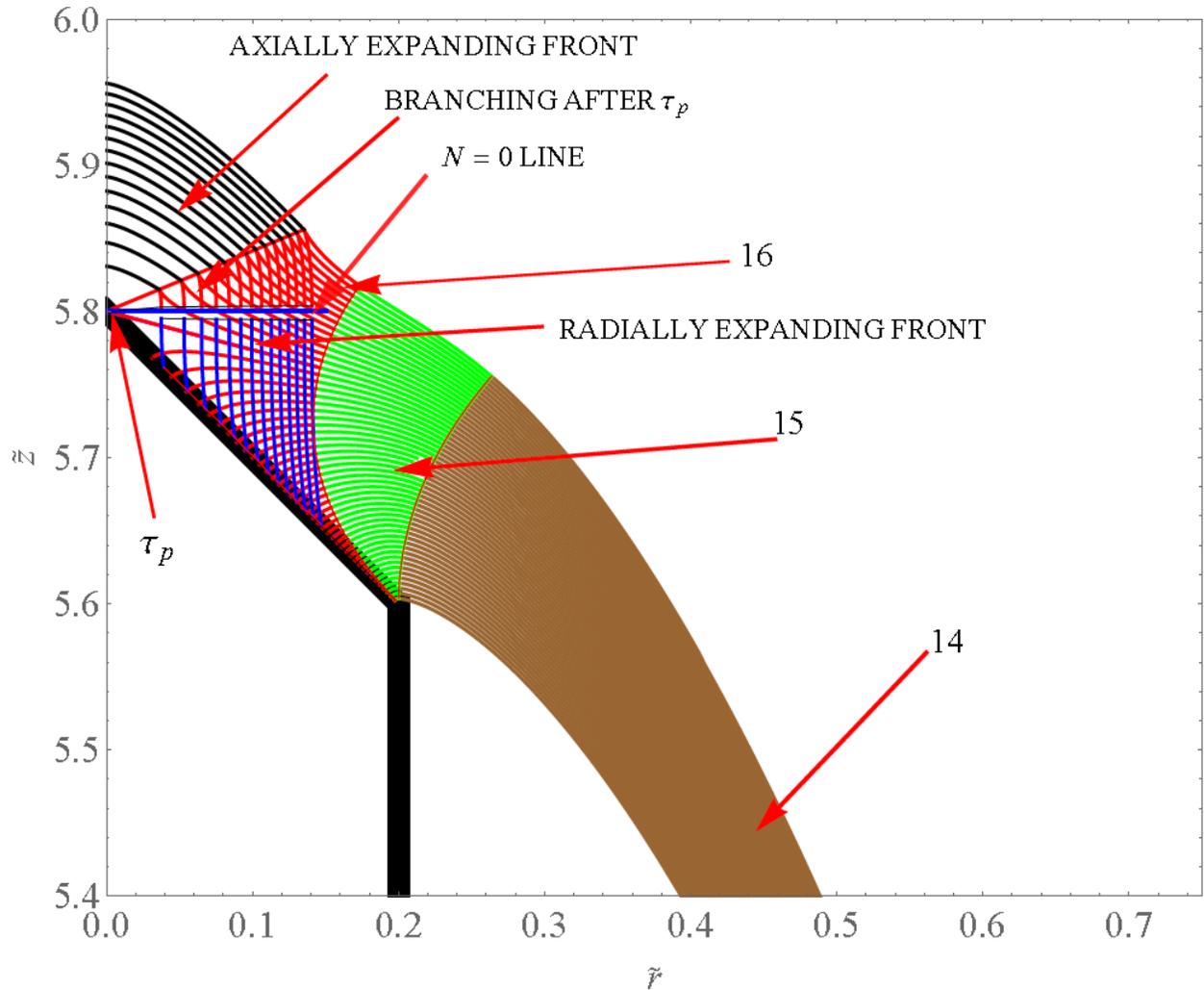

Fig 5: The GV surface in sub-region 16 travels along the tapered anode profile until it reaches the axis at $\tau_p$. For $\tau > \tau_p$, it undergoes branching. The locus of branch points is determined empirically and is shown by a red line starting from the anode tip. The branch points are joined to the Axially Expanding Front (AEF). The lower branch of the Radially Expanding Front (REF) ends at the N=0 line. It is continued towards the anode surface by the ghost element with negative sign of N in (47).

The noteworthy aspect of the figure is the transition of the GV surface, after it reaches the anode tip at $\tau_p$, into a qualitatively different curve with two branches with the same numerical procedure. The lower branch terminates on the N=0 line and the ghost segment of anode profile with negative sign of N in equation (47) connects it with the tapered anode. The space between the locus of branch points (connected by a line emanating from the anode tip) and the axis is filled by the Axially Expanding Front described by (77). Algorithmically, calculated points can be ascribed to the REF and CN branches depending on the sign of α.

A practical implementation of a type II profile can in principle be treated as a type I profile since a practically achievable anode tip can always be regarded as a flat top with a non-zero radius. The above illustration therefore serves mainly as a tutorial guide to the formalism.

However, one can think of an experiment where the straight segment A3 is in fact a hollow thin-walled metal tube (like a hypodermic needle), filled with deuterium gas at a high pressure and sealed. Then this anode profile serves as a plasma flow switch that transfers the current carried by the sheath to the sealed capsule that experiences a rapidly rising current that can launch an electrical explosion wavefront of the tube travelling axially along its outer wall and also launch a strong cone-shaped imploding shock wave into the gas, that is driven by a pinching magnetic field. Such shock wave would also be represented by the GV equation (28) and illustrates the point that the derivation of this equation does not need even the existence of the plasma focus electrode geometry.

**b. Illustration of a practical Mather type geometry with rounded edges and central cavity**

Real Mather type plasma focus devices have a central cavity and rounded edges. This profile can be represented as

$$\begin{aligned}
\tilde{R}_A(\tilde{Z}) &= 1 & \tilde{Z}_I \leq \tilde{Z} \leq \tilde{Z}_2 = \tilde{z}_A - \tilde{r}_1 & \Leftrightarrow A_1^{ext} \\
&= (1-\tilde{r}_1) + \sqrt{\tilde{r}_1^2 - (\tilde{Z}-(\tilde{z}_A - \tilde{r}_1))^2} & \tilde{z}_A - \tilde{r}_1 \leq \tilde{Z} \leq \tilde{z}_A & \Leftrightarrow A_2^{ext} \\
&= \tilde{r}_h + \tilde{r}_2 + \frac{(\tilde{Z}-\tilde{z}_{A+})}{(\tilde{z}_{A+}-\tilde{z}_A)}\left((\tilde{r}_h+\tilde{r}_2)-(1-\tilde{r}_1)\right) & \tilde{z}_A \leq \tilde{Z} \leq \tilde{z}_{A+} = \tilde{z}_A + \delta; \delta \ll 1 & \Leftrightarrow A_3^{ext} \\
&\uparrow \text{External branch} & & (83) \\
&= (\tilde{r}_h + \tilde{r}_2) - \sqrt{\tilde{r}_2^2 - (\tilde{Z}-(\tilde{z}_{A+}-\tilde{r}_2))^2} & \tilde{z}_{A+} - \tilde{r}_2 \leq \tilde{Z} \leq \tilde{z}_{A+} & \Leftrightarrow A_1^{int} \\
&= \tilde{r}_h & \tilde{z}_{A+} - \tilde{z}_h \leq \tilde{Z} \leq \tilde{z}_{A+} - \tilde{r}_2 & \Leftrightarrow A_2^{int} \\
&= \tilde{r}_h + \tilde{r}_h\left(\tilde{Z}-(\tilde{z}_{A+}-\tilde{z}_h)\right)/(\tilde{z}_{h+}-\tilde{z}_h) & \tilde{z}_{A+} - \tilde{z}_{h+} \leq \tilde{Z} \leq \tilde{z}_{A+} - \tilde{z}_h & \Leftrightarrow A_3^{int} \\
&\uparrow \text{Internal branch}
\end{aligned}$$

The insulator profile is assumed to be

$$\begin{aligned}
\tilde{R}_I(\tilde{Z}) &= \tilde{R}_I^{(0)} + \left(\tilde{R}_I^{(1)} - \tilde{R}_I^{(0)}\right)\tilde{Z}/\tilde{Z}_I^{(1)} & 0 \leq \tilde{Z} \leq \tilde{Z}_I^{(1)} & \Leftrightarrow I_1 \\
&= \tilde{R}_I^{(1)} + \left(\tilde{Z}-\tilde{Z}_I^{(1)}\right)\left(\tilde{R}_I^{(2)}-\tilde{R}_I^{(1)}\right)/\left(\tilde{Z}_I^{(2)}-\tilde{Z}_I^{(1)}\right) & \tilde{Z}_I^{(1)} \leq \tilde{Z} \leq \tilde{Z}_I^{(2)} & \Leftrightarrow I_2
\end{aligned} \quad (84)$$

It is easily seen that the central cavity (hole) has a normalized radius $\tilde{r}_h$ and the normalized fillet radius is $\tilde{r}_1$ at the outer edge of the anode and $\tilde{r}_2$ at the edge of the hole. This profile and the calculation of GV surfaces is illustrated in Fig 6. The plasma focus parameters are chosen for convenience of graphical illustration.

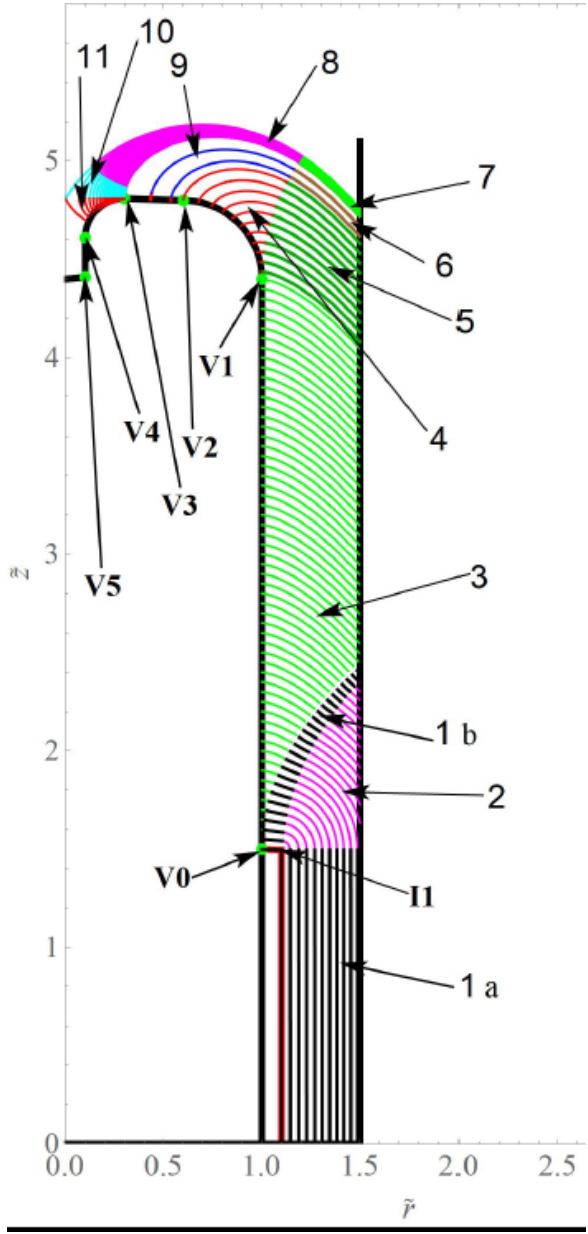

Fig 6: GV surfaces as a function of $\tau \leq \tau_p$. Device profile described by (83) and (84) shown in thick black lines. The parameters are as follows
$\tilde{R}_I^{(0)} = 1.1$, $\tilde{R}_I^{(1)} = \tilde{R}_I^{(0)} - 10^{-4}$,
$\tilde{Z}_A^{(0)} = \tilde{Z}_I^{(2)} = \tilde{z}_I = 1.5$, $\tilde{Z}_I^{(1)} = \tilde{Z}_I^{(2)} - 0.4 \times 10^{-4}$,
$\tilde{R}_A^{(1)} = \tilde{R}_I^{(2)} = 1$; $\tilde{Z}_2 = \tilde{z}_A - \tilde{r}_1$,
$\tilde{r}_1 = 0.4$, $\tilde{r}_2 = 0.2$, $\delta = 0.01$, $\tilde{r}_C = 1.5$
$\tilde{r}_h = 0.1$. The vertices of the insulator profile are (1.1,0), (1.0999,1.496), (1,1.5). The vertices of the external branch of the anode profile are
V0 (1, 1.5), V1 (1, 4.4), V2 (0.6,4.8), V3 (0.3, 4.81) and of the internal branch are
V3(0.3,4.81), V4 (0.1,4.61), V5 (0.1,4.41).
They are represented by green solid circles on the thick black anode profile.
The numerical values of parameters are chosen with the consideration of visual clarity of the illustration.

The times for the GV surface to reach these vertices starting from $\tau = 0$ at V0 are: $\tau_1 = 5.8$, $\tau_2 = 6.87398$, $\tau_3 = 7.14413$. The time to reach the axis is $\tau_p = \tau_3 + \tilde{R}_3^2 = 7.23413$. Times to reach the internal branch vertices are $\tau_4 = 7.25263$, $\tau_5 = 7.29263$ and $\tau_6 = 7.30268$. Clearly, the GV surface reaches the axis before it reaches the internal vertex V4. The GV surfaces are composed of 11 sub-regions, labeled with numerals and shown with different colors. They are shown in intervals of $\Delta\tau = 0.1$ for the external branch (sub-regions 1-9) starting from the last vertex of each segment and in intervals of $\Delta\tau = 0.005$ for the internal branch (sub-regions 10 and 11). Table II summarizes the parameters that are used to calculate them. The $(\tilde{r}, \tilde{z})$ values are matched at the boundaries of the sub-regions and at the cathode.

Table II: Parameters for calculating GV surfaces in Fig 6.

| # | Segment /Vertex | $\tilde{Z}$ | N | $\tau$ | s |
|---|---|---|---|---|---|
| 1.a | Segment $I_1$ | $(0 \cdots \tilde{Z}_I^{(1)})$ | $N_I(\tilde{Z})$ | $(0 \cdots \tau_1)$ | 1 |
| 1.b | Segment I2 | $(\tilde{Z}_I^{(1)} \cdots \tilde{Z}_I^{(2)})$ | $N_I(\tilde{Z})$ | $(0 \cdots \tau_1)$ | 1 |
| 2. | Vertex I1 | $\tilde{Z}_I^{(1)}$ | $(N_I^-(\tilde{Z}_I^{(1)}) \cdots N_I^+(\tilde{Z}_I^{(2)}))$ | $(0 \cdots \tau_1)$ | 1 |
| 3. | Segment $A_1^{ext}$ | $(\tilde{Z}_A^{(0)} \cdots \tilde{Z}_A^{(1)})$ | $N_A(\tilde{Z})$ | $(0 \cdots \tau_1)$ | -1 |
| 4. | Segment $A_2^{ext}$ | $(\tilde{Z}(\tau) \cdots \tilde{Z}_A^{(2)})$ | $N_A(\tilde{Z})$ | $(\tau_1 \cdots \tau_2)$ | -1 |
| 5. | Segment $A_1^{ext}$ | $(\tilde{Z}_A^{(0)} \cdots \tilde{Z}_A^{(1)})$ | $N_A(\tilde{Z})$ | $(\tau_1 \cdots \tau_2)$ | -1 |
| 6. | Segment $A_1^{ext}$ | $(\tilde{Z}_A^{(0)} \cdots \tilde{Z}_A^{(1)})$ | $N_A(\tilde{Z})$ | $(\tau_2 \cdots \tau_3)$ | -1 |
| 7. | Segment $A_1^{ext}$ | $(\tilde{Z}_A^{(0)} \cdots \tilde{Z}_A^{(1)})$ | $N_A(\tilde{Z})$ | $(\tau_3 \cdots \tau_p)$ | -1 |
| 8. | Segment $A_2^{ext}$ | $(\tilde{Z}(\tau) \cdots \tilde{Z}_A^{(2)})$ | $N_A(\tilde{Z})$ | $(\tau_3 \cdots \tau_p)$ | -1 |
| 9. | Segment $A_2^{ext}$ | $(\tilde{Z}(\tau) \cdots \tilde{Z}_A^{(2)})$ | $N_A(\tilde{Z})$ | $(\tau_2 \cdots \tau_3)$ | -1 |
| 10. | Vertex V3 | $\tilde{Z}_A^{(3)}$ | $(N_A^+(\tilde{Z}_I^{(3)}) \cdots 0)$ | $(\tau_3 \cdots \tau_p)$ | -1 |
| 11. | Vertex V3 | $\tilde{Z}_A^{(3)}$ | $(0 \cdots N_A^-(\tilde{Z}_A^{(3)}))$ | $(\tau_3 \cdots \tau_p)$ | -1 |

Fig 6 illustrates the seamless propagation of the solution over the curved portions of the anode profile from the straight portion. This validates the above generalization of the GV model algorithm applied to a classical Mather type geometry.

Fig 7 illustrates the formation of REF and AEF for $\tau_p + 0.0002$, $\tau_p + 0.0015$, $\tau_p + 0.005$ and $\tau_p + 0.01$ in both external and internal branches of the anode profile. The last vertex V3 is used to calculate these fronts. The GV surface is connected to the internal branch of the anode inside the cavity. Its connection to the cathode shows evidence of complex numerical behavior as seen in the series expansion (71) in the portion shown by magenta color. The expanded figure also shows a feature present but not clearly seen in Fig 6, namely, the red colored short stretch that connects the anode profile with the blue colored GV surface marked 9, which is calculated from the segment $A_2^{ext}$. This portion of GV surface is calculated from vertex V2.

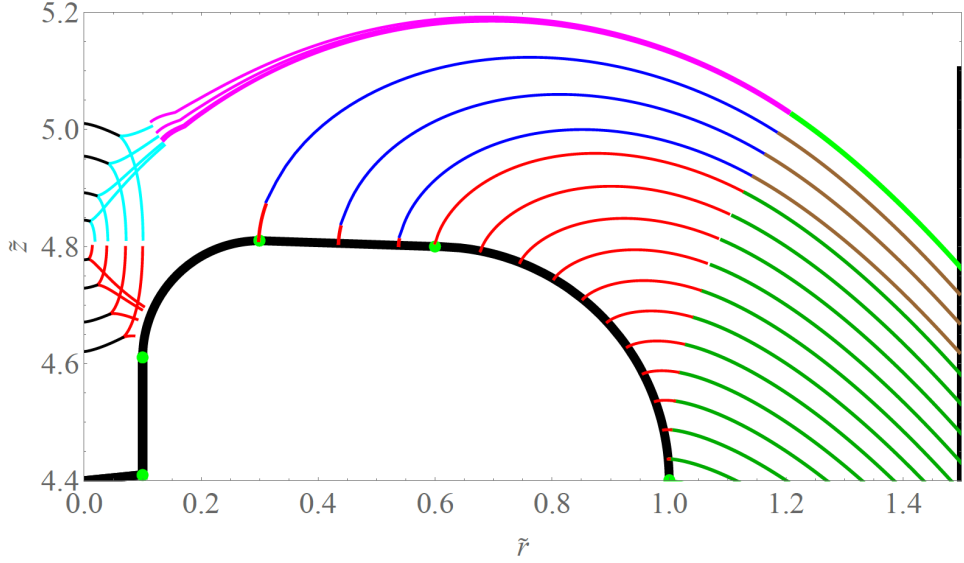

Fig 7: GV surface for $\tau > \tau_p$ for the same example as Fig 6. See text for details.

Note that the evolution of the GV surface near the axis and above the horizontal line joining the last vertex to the axis is similar to that of a conventional cylindrical anode without cavity.

## IV. Introducing a finite sheath thickness

The plasma focus sheath – which is usually understood to be the physical form of the plasma focus phenomenon between the discharge initiation phase and the pinch phase – is a complex formation whose detailed structure has not been experimentally investigated to the same degree of detail as the pinch phase. There are two main reasons for this. One is the doubly-curved shape of the sheath because of which plasma density, temperature, magnetic field, velocity all have a significant variation along its radial and tangential directions. Second is the limited number of diagnostic possibilities – visible light imaging, refractivity-based diagnostics and magnetic probes– because before the pinch phase, the plasma emits no x-rays /neutrons / fast ions /fast electrons that could serve as diagnostic probes.

The limited experimental information available for the sheath structure prior to the pinch phase may be summarized as follows

1. The sheath has a dense part with well-defined front and rear boundaries seen in refractivity and visible light emission (VLE) based diagnostics [2,19,26,27].

2. A low-density plasma trails behind the rear boundary whose density is very close to or even below the detection threshold of refractivity and VLE based diagnostics and whose presence is revealed only by magnetic probes [26,27,28] or by XUV imaging [29].

3. A multi-filamentary structure co-located with the sheath has been observed in VLE [30,31] and refractivity [32] based diagnostics.

4. Magnetic probe measurements [31] have reported a longitudinal component of magnetic field along the sheath that has been attributed to the magnetic structure of filaments.

5. Radial electron density profile of imploding sheath of Frascati 1 MJ plasma focus inferred from interferometry data [33,34] shows the following features

    a. The electron density rises to $1-2 \times 10^{17}$ cm$^{-3}$ over a distance of 3 mm at the leading front, which is numerically equal to the (atomic) number density of the fill gas at the operating pressure of 3 torr.

    b. After this, there is an abrupt rise of density to $1.6-1.8 \times 10^{18}$ cm$^{-3}$ over a distance of 3–4 mm followed by a steep decrease of density to a value below the detection limit over a scale of less than 1 mm.

    c. The peak electron density is about 2.5 times that indicated by Rankine-Hugoniot conditions for a planar steady shock.

6. Presence of axial (poloidal) magnetic field before the pinch phase is revealed by multiple diagnostics [12]. However, its relation with sheath structure is unknown at present.

7. Measurements using magnetic probes and 15-frame interferometry together have generated evidence that the current diffuses well inside the boundary of the dense plasma around the time of the current derivative singularity [4].

Theoretical efforts to understand these experimental observations [35,36,37,38] invoke radiative and collisional processes, making simplifying assumptions of a steady-state, planar plasma front. However, it is known that the magnetic pressure driving the plasma is not constant in time as it is dependent on the time-varying discharge current. Therefore, the physical situation is that of a non-steady non-planar shockwave. As can be demonstrated by simple considerations [39], a non-steady planar shock wave driven by an accelerating piston can be approximated by a sequence of steady planar shockwaves in time, which can reach effective compression ratios orders of magnitude higher than a single shock wave. Modeling efforts that ignore this [40,41] phenomenon and choose to assume results derived from the theory of steady planar shock waves run the risk of misrepresentation of the physical processes being modeled, jeopardizing conclusions derived from their extrapolations.

The present work adopts a different philosophy: it constructs a mathematical structure that *mimics* some aspects of the behavior of the plasma focus, with no claim to physical roots. It introduces a finite sheath thickness as a *kinematic artifact* rather than a physical phenomenon. The former could be introduced as a mathematical structure in a kinematic formalism with no pre-determined connection with any physical phenomenon while the latter would require formulation of a theory that takes into account interplay of many well-established physical phenomena [35-38] (atomic physics of collisions, ionization and radiation, conservation laws, transport of heat and electricity etc.).

Obviously, such approach cannot be a substitute for a physical theory of sheath structure. Accordingly, its purpose and scope need to be defined in clear terms. The purpose is to approximate the sheath in terms of a front boundary and a rear boundary which can be mathematically represented by imaginary surfaces of revolution in 3-dimensional cylindrical coordinates using the kinematic formalism developed in Section II. The scope is to examine which, if any, aspects of plasma focus phenomenology can be *mimicked* using this construction. Such aspects could then turn out to be insensitive to the microscopic details of physical phenomena known to be present in the plasma focus, providing important insights for the construction of a physical theory.

It should be specifically noted that, the *present work does not claim to be a theoretical representation of the plasma focus physics* [41] in its present stage of development and any resemblance of its conclusions with observed phenomenology must be treated as an independent new physical datum, that must be explained by a first principles theory, as and when one is developed.

Numerical simulations [18] relate the existence of a front boundary and a rear boundary to a shockwave driven into neutral gas and a magnetic piston driving it respectively. The rear boundary marks the region of the plasma focus electrode structure that contains the magnetic flux generated by the current flowing in the sheath at any time prior to stagnation and pinch formation and thus plays a significant role in the circuit element representation [19] of the plasma focus for a larger part of the propagation of the PCS. According to the heuristic slug model of Potter [40], the front boundary gets reflected from the axis and collides with the incoming rear boundary, stopping or slowing down its inward motion, *creating the initial boundaries of the z-pinch configuration* [40]. Experimental results indicate that *the current* completely relocates from the rear side of the magnetic piston into the boundary of the dense pinch. However, the bulk of the *magnetic flux* remains outside the boundary of the pinch.

The main motivation for introducing a finite thickness in the kinematic model is to enable it to mimic the finite size of the pinch using Potter's approach. The GV formalism is used to define a front boundary and a rear boundary and to mimic the reflection of the front boundary from the axis and its collision with the rear boundary. There is no intention of providing a physical theory that accounts for the observed sheath thickness and its variation in space and time in this Part I. The point about using a minimalist kinematic approach to such a complex situation is to provide a reference for quantifying the discrepancy between a simple model and experimental data that should point the way towards further refinement of the model into a closer approximation to a physical theory. This point is illustrated in Section V.

Although the kinematic framework of the model developed in Section II is versatile enough to calculate the GV surface for more general anode and insulator profiles, subsequent discussion in this section deals mainly with the classical Mather geometry (perhaps with a cavity and rounded edges) for the simple reason that much still remains to be understood about this reference configuration. Extrapolation of these results to a more general case needs to be attempted after a certain level of experimental corroboration is established regarding the present discussion involving the Mather geometry. The discussion below attempts to use nomenclature that would facilitate such future extrapolation.

Finite thickness is introduced in the kinematic model using the following two assumptions: (1) The rear boundary of the dense sheath corresponding to Potter's magnetic piston [40] is represented by the set of points $\left(\tilde{r}_{mp}(N,\tau), \tilde{z}_{mp}(N,\tau)\right)$ on the GV surface, parameterized by N, corresponding to $\chi = 1$ at dimensionless time $\tau$; this shall be often referred as the magnetic piston (MP) corresponding to Potter's model (2) the front boundary corresponding to Potter's hydrodynamic shock wave (SW) is represented by the set of points $\left(\tilde{r}_{SW}(N,\chi\tau), \tilde{z}_{SW}(N,\chi\tau)\right)$ on the GV surface corresponding to $\chi \gtrsim 1$ where $\chi$ is simply a model parameter for the present. The local normal velocity of the SW is $\chi$ times the local normal velocity of the MP as can be easily seen from the defining equation (20). Note that the use of the descriptors MP and SW is made only to refer to Potter's original hypothesis [40] about formation of the z-pinch boundary and does not imply incorporation of any additional physics used by Potter to compute these surfaces.

Fig. 8 illustrates the calculation of the two surfaces.

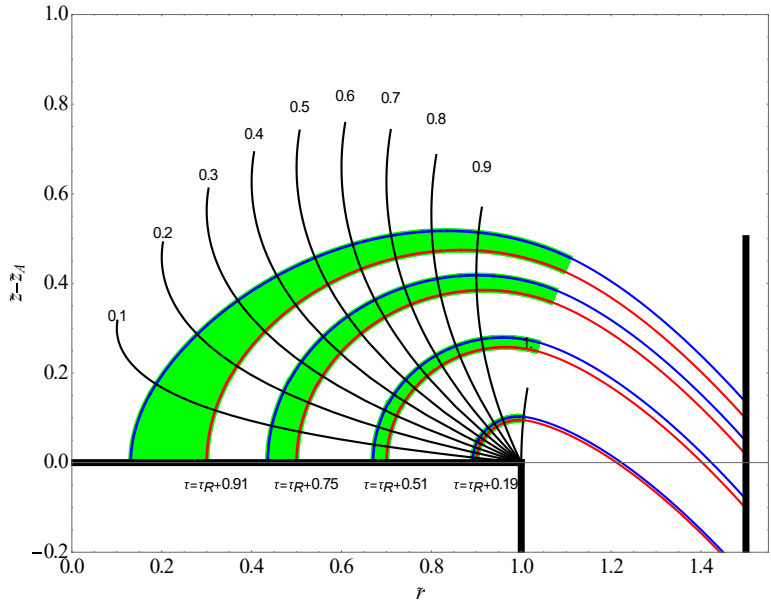

Fig 8: Illustration of finite thickness in the kinematic model for $\chi = 1.08$. The front of the dense sheath corresponding to the shock wave (SW) is shown with a thick blue line and the back corresponding to the magnetic piston (MP) with a thick red line. The black lines are characteristic curves emitted from the vertex $(1, \tilde{z}_A)$ at the end of the anode, labeled by values of N, which are the local normals to both the fronts. The shaded portions correspond to the GV surface normal to these characteristics and the un-shaded portion is part of the rundown phase solution extended beyond the rundown phase at $\tau = \tau_R$

The significance of the kinematic simulation of sheath thickness is seen as one advances the front of the sheath to the dimensionless time approaching $\tau_p$ – the value of $\tau$ when the solution of the GV equation reaches the axis. The SW surface forms a reflected solution that moves radially outward (REF-SW) and an axially moving solution (AEF-SW) that joins with the radially expanding branch to define an enclosed volume that expands monotonically (see Fig 3). A third branch (CN-SW) connects the junction of the two with the cathode and continues to move along its local normal with a radial velocity directed towards the axis. The MP continues to advance towards the axis even as the SW is expanding radially to intersect with it.

Taking cue from Potter's slug model [40] and the Lee model [41], the present model assumes that the rear boundary of the PCS stops its inward motion at the point of intersection between inward moving MP and outward moving SW. The boundary defined by the locus of Points Of Intersection (referred as POI) between the SW and the MP is identified as the zero-velocity boundary of the pinch in Potter's model.

In terms of 2-dimensional MHD simulations, this may be interpreted as follows: Symmetry about the axis dictates the boundary condition that the radial component $v_r$ of the velocity should be zero at the axis. Consequently, an outward moving fluid element is generated when the front of the plasma reaches the axis, which continues to move radially outwards. When this element reaches the outer boundary of the plasma, which is still moving inwards, the net fluid velocity becomes zero.

The transport of current in the sheath during its motion is a combined effect of convection and diffusion of magnetic field. At the point of intersection, the convection abruptly stops but the diffusion does not. Hence, the current diffuses into the zero-velocity boundary, as observed experimentally.

The task now is to determine the zero-velocity boundary of the plasma using the kinematic representation developed in Section II. This should be the locus of points of intersection $\left(\tilde{r}_{int}, \tilde{z}_{int}\right)$ between the outward-moving reflected SW and the still inward-moving MP occurring at $\tau_{int}$. From equation (74) above, the radius of the radially expanding SW at the anode surface is given by

$$\tilde{r}_{SW}^{2} = \chi(\tau - \tau_{last}) - \tilde{r}_{last}^{2} \tag{85}$$

For a simple Mather geometry, $\tilde{r}_{last} = 1$ is the anode shoulder and $\tau_{last} = \tau_R = 2(\tilde{z}_A - \tilde{z}_I)$ is the rundown time.

Similarly, the radius of the radially imploding MP at the anode surface is given by

$$\tilde{r}_{MP}^{2} = \tilde{r}_{last}^{2} - (\tau - \tau_{last}) \tag{86}$$

The SW and MP intersect at the anode surface at

$$\tau_{int} = \tau_p - \vartheta, \; \tilde{r}_{int} = \sqrt{\vartheta} \qquad \vartheta \equiv \tilde{r}_{last}^{2} \cdot (\chi - 1)/(\chi + 1) \tag{87}$$

From this instant onwards, they continue intersecting at distances further from the anode surface (see Fig 9) until at $\tau_p = \tau_{last} + \tilde{r}_{last}^{2}$, which equals $\tau_R + 1$ for a simple Mather geometry, the MP reaches the center of the anode and itself starts getting reflected. The radius of the reflected radially expanding front of the MP (REF-MP) at anode surface is given by

$$\tilde{r}_{MP}^{2} = (\tau - \tau_{last}) - \tilde{r}_{last}^{2} = \tau - \tau_p \tag{88}$$

which reaches $\tilde{r}_{int}$ at

$$\tau_E = \tau_p + \vartheta \tag{89}$$

From $\tau_p$ to $\tau_E$, the outward-moving REF-SW front intersects with the inward moving CN portion of the MP front, adding the point of intersection to the zero-velocity boundary of the pinch.

Since the MP and SW are moving mathematical surfaces rather than representations of fluid elements in the plasma, their motion governed by the GV equation does not stop at the point of intersection. It simply adds a point to the mathematical definition of the boundary.

The point $P_0(\tilde{r}_{int}, \tilde{z}_A)$ on the intersection of GV (MP) and $\chi$ (SW) surfaces with the anode forms the base of the boundary of the z-pinch "equilibrium" lying on the anode surface. This point remains stationary over the time interval $(\tau_{int}, \tau_E)$ according to the model assumption based on Potter's hypothesis since this is the time required for the GV (MP) surface to move from $\left(\tilde{r}_{int}, \tilde{z}_A\right)$ to the axis and back. The portion of the MP surface in the cavity shown in Fig 6 at $\tau_{int}$ would also remain stationary over the period $\left(\tau_{int}, \tau_E\right)$, connecting the zero-velocity boundary with the cavity inside the anode in view of the transformation properties of (41) discussed above.

The disposition of the GV (MP) and $\chi$ (SW) surfaces at the three times $\tau_{int}$, $\tau_p$ and $\tau_E$ is illustrated in Fig 9. The equilibrium boundary of the pinch at the time $\tau_E$ (shown as a thick green line) is composed of the points of intersection (POI) between incoming MP and outgoing SW-REF. At times $\tau$ between $\tau_{int}$ and $\tau_E$, the equilibrium boundary consists of the collection of points of intersection up to $\tau$ and the MP at $\tau$.

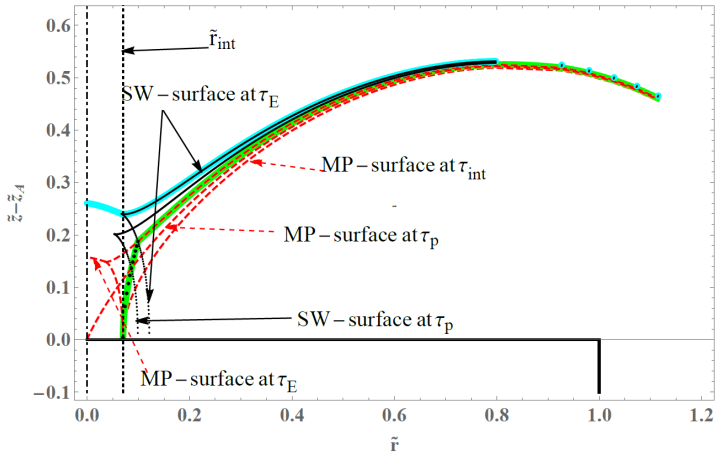

Fig. 9: The thick green line represents the boundary of the "equilibrium pinch". Black dots overlaid on the green line are the Points of Intersection (POI) between the reflected Radially Expanding Front of the SW surface and the inward moving portion of the MP surface at time intervals $\vartheta/5$. Three positions of the MP surface at $\tau_{int}$, $\tau_p$ and $\tau_E$ are shown using dashed red lines. The SW surfaces at $\tau_p$ and $\tau_E$ are shown using black dots. The position of the SW surface at $\tau_{int}$ is the same as that of MP surface at $\tau_E$. The thick Cyan line is the external boundary of the pinch equilibrium. It is composed of the Axially Expanding Front and the connector branch of the SW surface at $\tau_E$. This and subsequent figures use $\chi = 1.01$ for which $\vartheta = 0.00402985$.

The evolution of the pinch configuration according to this model is illustrated in Fig 10, 11 and 12.

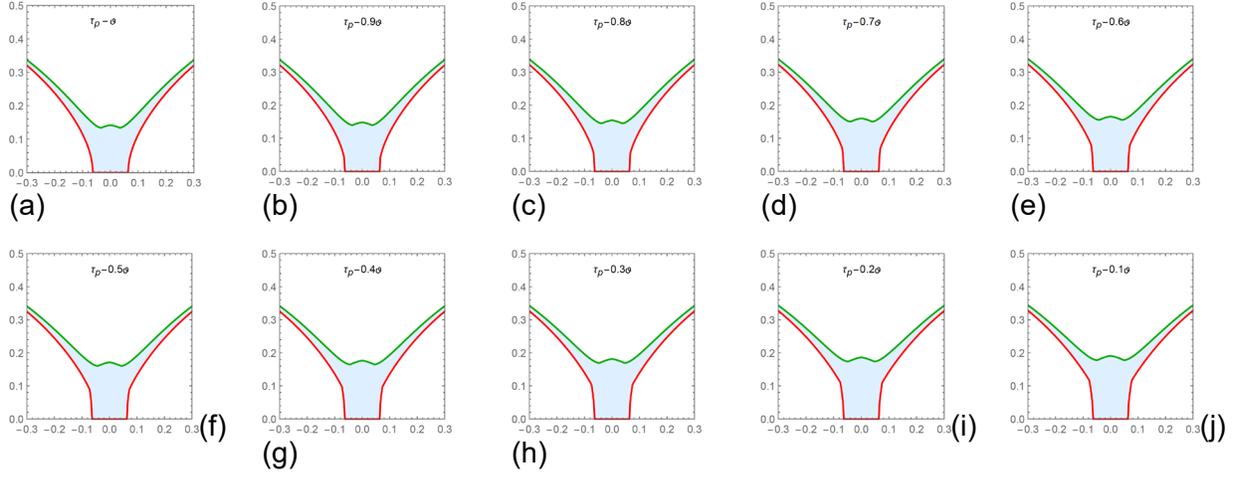

Fig 10: Evolution of equilibrium pinch boundary from $\tau_{int} = \tau_p - \vartheta$ to $\tau_p$. In this period, the SW front is divided into a Radially Expanding Front (REF), an Axially Expanding Front (AEF) and a connector (CN), while the MP front has yet to reach the axis at $\tau_p$. The upper portion of the boundary is formed by joining AEF-SW and CN-SW. The lower portion is made by joining the locus of the Points of Intersection (POI) between the REF-SW and MP with the portion of CN-MP. The sequence of figures shows how the "stem" portion develops.

It is noteworthy that the GV (MP) surface also generates an Axially Expanding Front (AEF-MP) after $\tau_p$ which provides a sort of rear boundary to the Axially Expanding Front generated by the SW surface formed at times later than $\tau_{int}$. Such bounded axially moving structures are experimentally observed (see Fig 6 of Ref [42]). An analog of this phenomenon can be constructed in this model by joining pieces of the MP and SW surfaces that are not used in constructing the upper and lower surfaces: the AEF-MP, portion of the CN-MP up to its intersection with REF-SW, the portion of REF-SW beyond this intersection up to its intersection with REF-CN. From this intersection, the AEF-SW forms the upper portion of the pinch, but its mirror image can be used up to the axis to form the closed boundary of a bounded structure. Fig 11 shows this phenomenon occurring from $\tau_p$ to $\tau_E = \tau_p + \vartheta$.

The intersections of the axis with the top and the bottom of the bounded structure has the following regression fits

$$\tilde{z}_{top} = 0.173087 + 6.00647\,(\tau - \tau_p);\ \tilde{z}_{bottom} = 1.79241(\tau - \tau_p)^{0.459982};\ 0 < \tau - \tau_p < 2.5\vartheta \qquad (90)$$

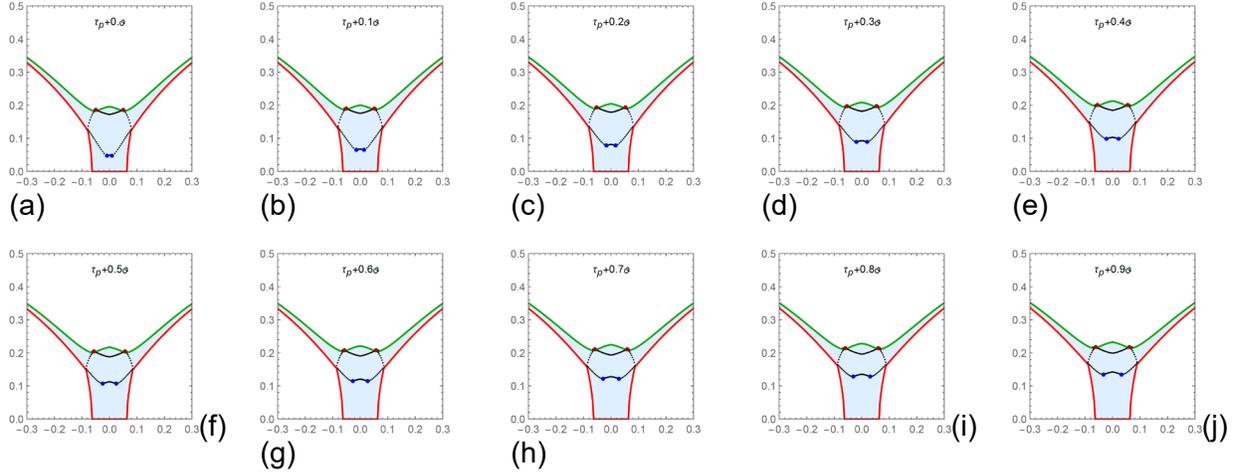

Fig 11: Evolution of equilibrium pinch boundary from $\tau_p$ to $\tau_E = \tau_p + \vartheta$. In this period, both the SW and MP fronts are divided into a Radially Expanding Front (REF), an Axially Expanding Front (AEF) and a connector (CN). The upper portion of the boundary is formed by joining AEF-SW and CN-SW. The lower portion is made by joining the locus of the Points of Intersection (POI) between the REF-SW and CN-MP with the remaining portion of CN-MP. The black dotted contour is made from (1) AEF-MP at the bottom, (2) portion of the CN-MP between its junction with the AEF-MP (blue dot) and POI between REF-SW and CN-MP (3) portion of REF-SW between its junction with AEF-SW (red dot) and POI between REF-SW and CN-MP (4) mirror image of AEF-SW. Its shape resembles the contour map of the squared-modulus of an m=0 Chandrasekhar-Kendal function [43]

The noteworthy aspect of this exercise is the marked similarity between the shape of the resulting bounded structure and the shape of the contour map of the squared-modulus of the m=0 Chandrasekhar-Kendal function [43]. This could in principle be used to construct a theory of a Turner Relaxed State [43] defined within this bounded structure. It would be worth noting at this stage that toroidal structures similar to the m=0 curl eigenfunctions have been observed to be correlated with the neutron emission [4].

After $\tau_E$, the base of the pinch boundary again starts moving outwards with the local velocity of the radially expanding MP front. From this instant onwards, the REF-MP starts intersecting with the zero-velocity boundary of the "equilibrium pinch" represented by the POI which gradually gets "deformed". The boundary of the z-pinch after $\tau_E$ can then be constructed from the portion of the REF-MP between the anode and this point of intersection, the remaining portion of the equilibrium boundary and the portion of the CN-MP branch to the higher radius side of its point of intersection with the REF-SW. This leads to boundaries that have a higher cross-section near the anode than near the top of the umbrella shape. The gradual evolution of this shape is illustrated in Fig 12. Similar shapes are found in interferometric images from PF-1000 [44] and POSEIDON [45].

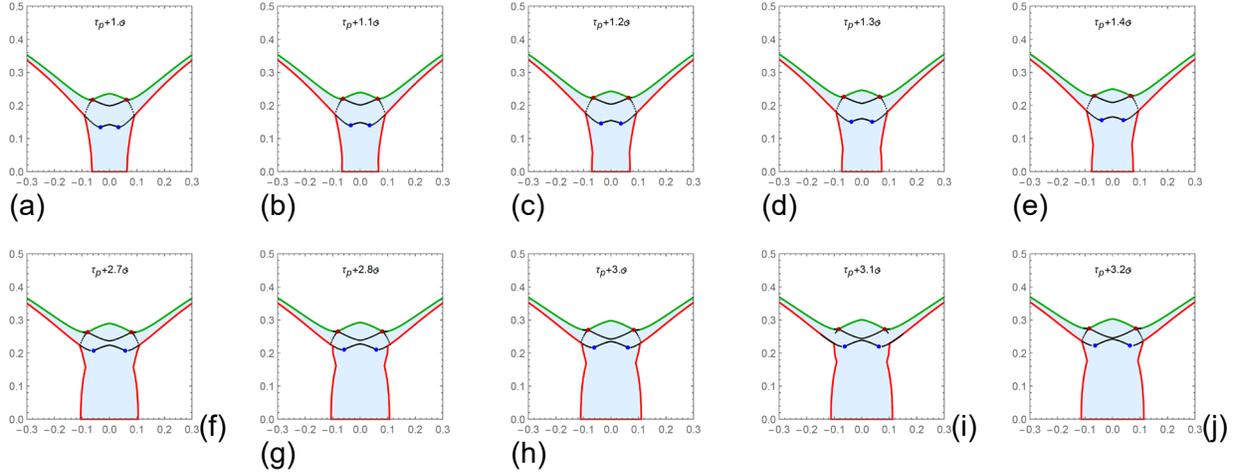

Fig 12: Evolution of the pinch boundary beyond $\tau_E = \tau_p + 9$. This is the time when REF-MP reaches the base of the POI that form the equilibrium pinch boundary (see Fig 9) and corresponds to (a). The upper portion of the boundary is formed by joining AEF-SW and CN-SW as in Fig 11. See text for description of the lower boundary. Note how the embedded bounded shape is getting axially squeezed.

As mentioned earlier, $\chi$ is a model parameter that introduces a finite sheath thickness as a kinematic artifact that has *no relation to physical reality unless (and only to the extent) the model is used for comparison of its predictions with experimental measurements*. The following example illustrates this assertion.

The profile of Fig 12(a) is at $\tau_E = \tau_p + 9$, which is of special significance as having the largest pinch height before the onset of pinch expansion that begins near the anode. It can be distinguished visually from the next profile (b) by the absence of the noticeable bulge in diameter near the anode.

Fig 12(a) can be compared Fig 2(a) (at 3 ns) of Ref 44 where the plasma radius is 8 mm and height is 30 mm. Its compact size as compared with its preceding (-27 ns) and succeeding (33 ns) images, its temporal closeness (3 ns) to the current derivative minimum and the noticeable axial taper suggest that it corresponds to the time $\tau_E$. With the anode radius of 115 mm, normalized radius and height of the pinch are 0.07 and 0.26, respectively. The regression fit (79) for the position of the AEF in a classical Mather type plasma focus gives the following scaling relation for the position of the AEF at $\tau_E = \tau_p + 9$

$$\tilde{z}_{APF}(\tau_E) = 1.70246 \left( \frac{2\chi^2}{\chi+1} - 1 \right)^{0.443703} \tag{91}$$

which yields $\chi = 1.00964$ assuming that this corresponds with observed normalized pinch height ~0.26 [44]. For this value of $\chi$, formula (87) gives $\tilde{r}_{int} \sim 0.0695$, in surprisingly good agreement with the experimental normalized radius of 0.07. This corresponds almost exactly with Fig 12(a), drawn assuming $\chi = 1.01$. Another example is Fig 2 of Ref 45, which depicts Abel inversion of the pinch phase of POSEIDON for shot #6308. The radius of the plasma stem is ~8 mm and the height is ~28 mm. The anode radius is 65.5 mm. The normalized plasma height is ~0.427, while the normalized plasma radius is ~0.12. For this value of normalized height, (91) gives $\chi = 1.02938$ for which (87) gives $\tilde{r}_{int} \sim 0.120328$, again a very good agreement. Note that the scaling model proposed by Lee and Serban [46] provides the values of scaled radius and height as 0.12 and 0.8 for all plasma focus devices.

This observation allows the model to *choose to interpret* the enclosed volume surrounding the axis in Fig 10,11 and 12 as the dense plasma pinch that is visible in interferometry [44,45], schlieren or x-ray pictures. The apparent similarity between the "cartoon" seen in Fig 9,10,11 and 12 and the plasma shape seen in schlieren diagnostics [11] and interferometry [44] suggests that the model can mimic the umbrella shape of the pinch phase. Fig 13 shows a surface of revolution constructed from the profile of Fig 12(a).

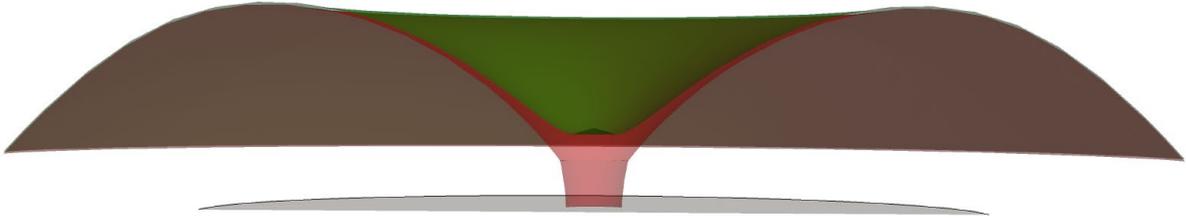

Fig 13: Surface of revolution constructed from the profile viewed in cross-section with reduced opacity.

The definition of the parameter $\chi$ in equation (21) suggests that the local normal velocity of the SW front is $\chi$ times the local normal velocity of the MP front. One can *choose to interpret* the pair of local points on SW and MP for the same value of $\tau$ and N as *simultaneously belonging to a planar steady shock wave driven by a planar piston moving with constant velocity*. This cannot be rigorously justified by a "first principles" physical theory and must be treated as an ad hoc construction in the spirit of the slug model [40] and the Lee model [41]. From the theory of high Mach number planar steady shock waves driven by a piston, the following relations [40] can be borrowed

$$v_{shock} = \tfrac{1}{2}(\gamma+1)u_{piston}; \frac{\rho_{shock}}{\rho_0} = \left(\frac{\gamma+1}{\gamma-1}\right) \tag{92}$$

in order to define a fictitious "kinematic adiabatic index"

$$\gamma_k \equiv 2\chi - 1 \tag{93}$$

and a "kinematic density behind the shock"

$$\rho_k \equiv \rho_0 \left(\frac{\gamma_k+1}{\gamma_k-1}\right) \tag{94}$$

For the PF-1000 example, $\gamma_k = 1.01927$, giving $\rho_k \approx 105\rho_0$. At 100 Pa filling pressure at 20°C[44], the atomic number density of D2 gas is $\sim 4.94 \times 10^{16} \text{cm}^{-3}$ so the kinematic electron number density from (94) $n_k \sim 5.2 \times 10^{18} \text{cm}^{-3}$ which agrees with the peak electron density in the pinch at 3 ns in Fig 2(a) of [44] within 15%. For POSEIDON, $\gamma_k = 1.05877$, giving $\rho_k \approx 35\rho_0$. The filling pressure of 5 hPa of D2 at 20°C corresponds to an atomic number density of $\sim 2.47 \times 10^{17} \text{cm}^{-3}$. This yields for POSEIDON, the electron number density $n_k \sim 8.64 \times 10^{18} \text{cm}^{-3}$ which agrees with the peak electron number density of Fig 2 of [45] within 15%.

Using the kinematic density and the computed umbrella shape, it is possible to generate a synthetic interferogram. Assuming that the laser beam travels along the x-direction, the phase shift recorded on the image plane parallel to the (y, z) plane would be

$$\varphi(y,z) = \frac{\pi a}{\lambda n_c} \int_{-1}^{1} n(\tilde{x},\tilde{y},\tilde{z})d\tilde{x} = 2.82 \times 10^{-21} \cdot a(m)\lambda(\mu m)\int_{-1}^{1} n(\tilde{x},\tilde{y},\tilde{z})d\tilde{x} \tag{95}$$

where $n_c$ is the critical plasma density for the laser wavelength. The kinematic density profile $n_k(\tilde{r},\tilde{z})$ can be described as

$$\frac{\rho_k}{m_i} \equiv 4.8286 \times 10^{20} \left( m^{-3} \right) \cdot p_{D_2} (Pa) \left( \frac{\gamma_k + 1}{\gamma_k - 1} \right) \tag{96}$$

multiplied by 1 when the point $(\tilde{r}, \tilde{z})$ is within the region bounded by the SW and the MP and zero otherwise. Thus

$$\varphi(\tilde{y}, \tilde{z}) = 1.3616652 \cdot p_{D_2} (Pa) \left( \frac{\gamma_k + 1}{\gamma_k - 1} \right) \cdot a(m) \lambda(\mu m) \int_{-1}^{1} \mathbb{B}(\tilde{x}, \tilde{y}, \tilde{z}) d\tilde{x}$$
$$\mathbb{B}(\tilde{x}, \tilde{y}, \tilde{z}) = 1 \quad \text{if } \tilde{r}_{SW}(N, \tau) \leq \sqrt{\tilde{x}^2 + \tilde{y}^2} \leq \tilde{r}_{mp}(N, \tau) \quad \text{and} \quad \tilde{z}_{SW}(N, \tau) \geq \tilde{z} \geq \tilde{z}_{mp}(N, \tau) \tag{97}$$
$$= 0 \quad \text{otherwise}$$

Note that this completely ignores the bounded structure depicted in Fig 11 and 12. For the PF-1000 example cited above [44], the fill pressure is 100 Pa, $\lambda(\mu m) = 0.527$, $a(m) = 0.115$. The following figure 14 is a numerically constructed image of $\sin^2 \varphi(\tilde{y}, \tilde{z})$.

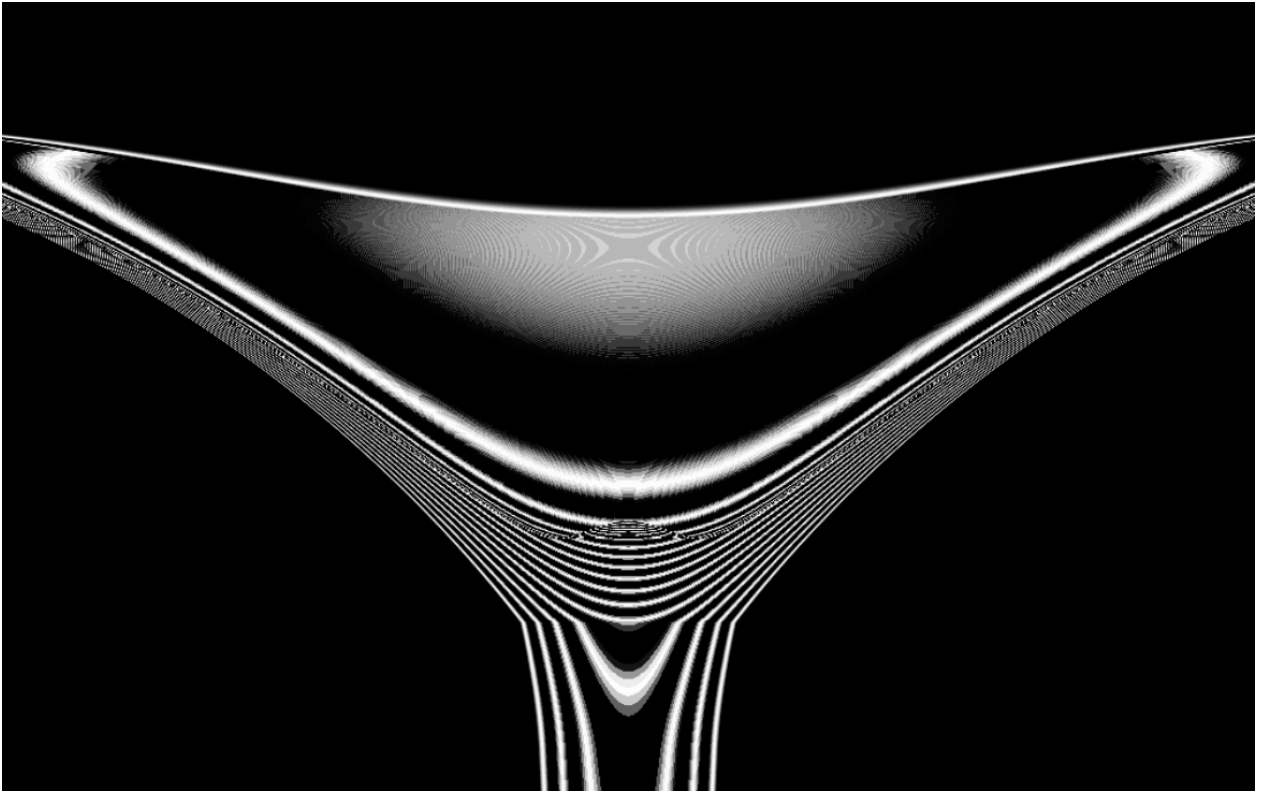

Fig 14: Numerically constructed image of $\sin^2 \varphi(\tilde{y}, \tilde{z})$ using Mathematica® Image function

The above discussion shows that the introduction of a finite sheath thickness as a kinematic artifact is versatile enough to *mimic* the scaled height and radius of the pinch stem, the umbrella shape and the density the pinch. The inward motion of the idealized rear boundary, its temporary halt and subsequent outward motion are similar to the observed streak pictures (see fig. 6 of [47]).

It also has a feature that resembles bounded plasmoidal / toroidal structures observed in the plasma focus [4] which are a major ingredient of the Rotating Fountain BEAM Plasmoid TARGET fusion mechanism proposed [4] to explain a large amount of the available data. This strongly suggests that these aspects of the plasma focus could turn out to be insensitive to details of physical phenomena occurring at the time of pinch formation – a suggestion that needs to be cross-checked against data from many experiments.

### V. Model closure and the circuit equation:

The considerations so far show that the kinematic framework is able to mimic many aspects of the plasma focus evolution as a function of the dimensionless time $\tau$. They also reveal that the kinematic model reaches its limits of utility once the zero-velocity boundary of the pinch phase is reached. This limit is represented by the time $\tau_E$, beyond which, more involved physical phenomena need to be taken into account. In order to obtain numbers that can be compared with measurements, which is the primary goal of a model, it is necessary to achieve model closure by recovering real time t from the dimensionless time $\tau \leq \tau_E$ by inverting (27) as

$$t = t_L + Q_m \int_0^\tau \frac{d\tau}{I(\tau)} \tag{98}$$

To facilitate further discussion, the following dimensionless quantities and scaling factors are defined

$$\alpha = 1 - Q_L/Q_0 = \exp(-\gamma_0 t_L)\left(\cos(\omega_0 t_L) + \frac{\gamma_0}{\omega_0}\sin(\omega_0 t_L)\right); \quad \varepsilon \equiv Q_m/C_0 V_0; \quad \kappa \equiv \mu_0 a/2\pi L_0;$$

$$I_0 \equiv V_0\sqrt{C_0/L_0}\,;\ \tilde{I}(\tau) \equiv I(\tau(t))/I_0\,;\ \gamma \equiv R_0\sqrt{C_0/L_0} = 2\lambda\,;\ T_{1/4} = \frac{\pi}{2}\sqrt{L_0 C_0} \tag{99}$$

The real time can be expressed in dimensionless form by scaling it with the ideal quarter cycle time:

$$\tilde{t} \equiv \frac{t}{T_{1/4}} = \frac{\omega_0 t_L}{\frac{1}{2}\pi\sqrt{1-\lambda^2}} + \frac{2\varepsilon}{\pi}\int_0^{\tilde{\tau}} \frac{d\tau}{\tilde{I}(\tau)} \tag{100}$$

According to the discussion of Section IIB, obtaining the value of $t_L$ requires the empirical determination of the lower velocity threshold $v_{LB}$. This can be done using two shots of the plasma focus. One is taken at the lowest reliable operating voltage and high gas density (including use of a heavy gas such as argon or using high pressure of deuterium). This should give a current signal having the shape of (14), perhaps with some offsets in both the time and the voltage depending on the actual oscilloscope settings. Fitting the current waveform to this formula should yield values of the circuit constants $L_0$ and $R_0$ as well as the calibration constant for the Rogowski coil + integrator. The second shot is taken at any pressure that produces a current derivative singularity. Integration of the current signal up to the time of the singularity then gives the charge $Q_p$ at $\tau_p$. Then the definition (27) of $\tau$ can be used to write

$$Q_L = Q_p - Q_m \tau_p \tag{101}$$

With a measured value of $Q_p$ and theoretical values of $Q_m$ and $\tau_p$, $Q_L$ is determined, from which (19) can yield the value of $t_L$. Then using (16), $v_{LB}$ can be calculated. This should be a material constant for the fill gas, so can be determined once and for all. Then for any other pressure or device with the same gas, $t_L$ can simply be calculated using the formalism of Section II B.

Model closure using (98) requires setting up and solving the circuit equation for the current $\tilde{I}(\tau)$. While Section IV limited the discussion to only Mather type devices, the model closure is a fundamental requirement of the general case. This section therefore reverts to the general case considered in Section II. However, an operational Mather type plasma focus [48] will be used for illustration.

The following discussion is based on a circuit schematic, shown in Fig. 15 which is divided into a power source and the plasma focus. The junction between the two is shown by the terminals labelled 1 and 2, with a voltage $V(t)$ across them. A current $I(t)$ flows clockwise in the circuit. Since there are no batteries, the total loop voltage in the direction of the current is zero.

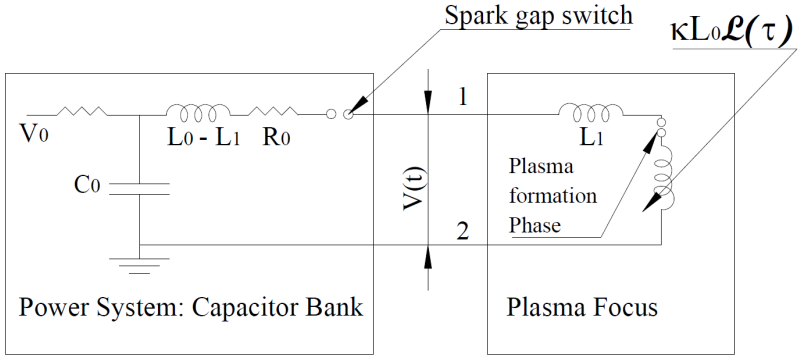

Fig. 15. Schematic of the circuit. The power system is shown as an LCR circuit with constant parameters $C_0$, $R_0$ and a part of the total static inductance of the circuit $L_0 - L_1$. The plasma focus is shown as an inductive element with static inductance $L_1$ and a time-varying inductance. The plasma formation phase is represented as a small spark gap switch in series with the inductance.

The circuit up to time $t_L$, at which the plasma begins its propagation, is the standard LCR circuit whose solution for current $I_s(t)$ is given in equation (14). The objective now is the determination of current evolution beyond this instant. This requires the representation of the plasma focus as a circuit element.

This question has been examined from first-principles, whose central point is best explained by the following quote [19]:

"…while representing the spatially extended electrodes as point terminals of an idealized circuit element, the voltage $V_{12}$ [across electrodes labeled 1 and 2] must be considered as the *average* work done while transporting the test charge between a pair of points on the two electrodes along the streamlines of electric field connecting the pair. This provides the following physical basis for constructing a circuit element representation:

$$V_{12}(t) = -\frac{1}{I(t)} \int_\Omega d^3\vec{r}\, \vec{J} \cdot \vec{E} \qquad (102)$$

where the right-hand side is the total electric power flowing through the two terminals expressed in terms of fields inside the device divided by the total current flowing through the device. *Every physical phenomenon, without exception*, that happens within the plasma chamber occurs because of the power and current supplied from outside. The expression for the voltage across the terminals in a circuit representation must allow the power required by each of such physical phenomena to be drawn from the electrical circuit in an adequate manner."

The discussion leads to the following expression for the product of voltage across and current through the connection of the plasma focus with the capacitor bank, where the 3-D domain $\Omega$ (see Fig 1 of Ref 19) is bounded by the electrode surfaces and the rear surface $\Sigma_p$ of the plasma sheath, (designated as MP in the present paper).

$$I(t)V(t) = \underbrace{\frac{d}{dt}\int_\Omega d^3\vec{r}\left(\tfrac{1}{2}\mu_0^{-1}B^2\right)}_{I} + \underbrace{\int_{\Sigma_p} d\vec{S}\cdot\vec{v}\left(\tfrac{1}{2}\mu_0^{-1}B^2\right)}_{II} + \underbrace{\frac{d}{dt}\int_\Omega d^3\vec{r}\left(\tfrac{1}{2}\varepsilon_0 E^2\right)}_{III} - \underbrace{\int_{\Sigma_p} d\vec{S}\cdot\vec{v}\left(\tfrac{1}{2}\varepsilon_0 E^2\right)}_{IV}$$
$$+\underbrace{\mu_0^{-1}\oint_{\Sigma_p} d\vec{S}\cdot\eta\vec{J}\times\vec{B}}_{V} - \underbrace{\mu_0^{-1}\oint_{\Sigma_p} d\vec{S}\cdot\vec{B}\left(\vec{B}\cdot\vec{v}\right)}_{VI} \quad (103)$$

In (103), $\vec{v}$ is the velocity of the rear surface of the plasma sheath. Note that the derivation of this expression neglects the Hall term and the electron pressure gradient terms in the Generalized Ohm's Law, which must be taken into account in a more complete theory.

This general expression needs to be adapted to the case of the present kinematic model. In low impedance systems such as the plasma focus, the energy density present in electric field is always much less than the energy density in the magnetic field. The third and fourth terms of (103) can therefore be neglected as compared to the first two terms during the propagation of the sheath. The fifth term proportional to the resistivity can be neglected because of high Reynold's number ( $R_M \sim 80$ for commonly accepted orders of experimental numbers $v_n \sim 10^5 m/s$, width of current distribution $\sim 2\times 10^{-2} m$, $\eta_{Spitzer} \sim 3.\times 10^{-5}\Omega - m$ assuming $T_e \sim 20eV$ ). The sixth term is neglected because the kinematic framework does not have parallel magnetic and velocity fields.

These adaptations underscore the fact that the kinematic framework is not expected to be a complete representation of plasma focus physics by itself. The circuit element representation needs to account for the recent observations [49] of poloidal magnetic flux emission before, during and after the pinch phase. This effect can be looked upon as a consequence of generation of an azimuthal current because of the motion of an azimuthally continuous, curved conductor in the geomagnetic field. Magnitude of this current should scale with the amplitude of the geomagnetic field, with the local instantaneous sheath velocity that scales as $I(t)/\tilde{r}$ and inversely as the plasma resistivity. This would produce a poloidal magnetic field $\vec{B}_p = B_r \hat{r} + B_z \hat{z}$ everywhere. It would also introduce an azimuthal velocity $v_\theta$ because of conservation of angular momentum. Treating the geomagnetic field as a small perturbation, the quantities $v_\theta$, $B_r$ and $B_z$ must be considered to be of first order in the perturbation while $B_\theta$, $v_r$ and $v_z$ are of zeroth order. However, dynamo effects that convert the kinetic energy of the plasma into magnetic energy associated with the poloidal magnetic field and momentum convection effects that transfer kinetic energy from radial to azimuthal motion are expected to amplify these first order effects as the sheath approaches the axis because of their inverse dependence on the plasma radius. This process depends on the spatial distribution of plasma velocity between the front and rear boundaries, which is outside the scope of the kinematic model. The bottom line is that the predictions of the circuit representation in the kinematic model are *expected* to deviate from the experimental data because of the diversion of electrical power supplied by the power source into physical phenomena not accounted for, representing a loss of current of unknown origin [50].

The voltage across the connection of the PF with the capacitor bank (or any other power source such as an inductive storage or an explosive flux compression device) is then given by

$$V(t) = \frac{1}{I(t)} \left( \frac{d}{dt} \int_\Omega d^3\vec{r} \left( \tfrac{1}{2} \mu_0^{-1} B^2 \right) + \int_{\Sigma_P} d\vec{S} \cdot \vec{v} \left( \tfrac{1}{2} \mu_0^{-1} B^2 \right) \right) \qquad (104)$$

Note that this is the voltage of terminal 1 with respect to terminal 2. While calculating the loop voltage, one follows the direction of the current. The contribution of the plasma focus as circuit element to the loop voltage will be $-V(t)$.

Using equation (25), (104) can be rewritten as

$$I(t) V(t) = \left\{ \frac{d}{dt} \left( \frac{\mu_0}{8\pi^2} I^2(t) \int_\Omega \frac{d^3\vec{r}}{r^2} \right) + \frac{\mu_0 I^2(t)}{8\pi^2} \int_{\Sigma_P} \frac{d\vec{S} \cdot \vec{v}}{r^2} \right\} \qquad (105)$$

From Reynold's Transport Theorem [51],

$$\frac{d}{dt}\int_\Omega \frac{d^3\vec{r}}{r^2} = \int_{\Sigma_p} \frac{d\vec{S}\cdot\vec{v}}{r^2} \tag{106}$$

Using dimensionless variables and (106), (105) can be written as

$$V(t) = \frac{\kappa}{\varepsilon} V_0 \left\{ \frac{d}{d\tau}\left(\frac{1}{2}\tilde{I}^2(\tau)\mathcal{L}(\tau)\right) + \frac{1}{2}\tilde{I}^2(\tau)\frac{d}{d\tau}\mathcal{L}(\tau) \right\} \tag{107}$$

where,

$$\mathcal{L}(\tau) \equiv \frac{1}{2\pi}\int_\Omega \frac{d^3\tilde{\vec{r}}}{\tilde{r}^2} = \int_\Omega \frac{d\tilde{r}d\tilde{z}}{\tilde{r}} \tag{108}$$

Equation (107) demonstrates that the circuit element representation of the plasma focus, under the conditions where the last 4 terms of (103), along with Hall effect and electron pressure gradient terms in the Generalized Ohm's Law, can be neglected, is indeed a time-varying inductance defined as flux per unit current:

$$L(t) \equiv \frac{1}{I(t)}\int B_\theta(r,t) dr dz = \frac{\mu_0}{2\pi}\int \frac{drdz}{r} = \kappa L_0 \mathcal{L}(\tau) \tag{109}$$

This also means that the inductance model must necessarily fail some stringent experimental tests. Two examples of such failures are known [19]. One is the increase in the inductance calculated from experimental current and voltage traces after the pinch phase. Since the plasma is known to expand radially [44, 45], its inductance should rather decrease. The second is the observation in a neon plasma focus of a pinch forming and breaking at least 200 ns before the current derivative singularity. This delay cannot be understood if the current derivative singularity is assumed to be a consequence of plasma inductance rise with decreasing plasma radius. As discussed above, the observation of emission of poloidal magnetic flux [49] indicates that the inductance model of the plasma focus needs corrections outside the kinematic framework taking into account relevant physical. Its main role is, therefore, to provide closure to the kinematic model within its stated objectives and limitations.

The plasma focus is known to be operated with inductive storage [52] and explosive pulse power generators [53,54], besides the conventional capacitor bank. In this paper, the power source is represented as an idealized capacitor bank with constant parameters $L_0$, $R_0$ and $C_0$. However, a similar procedure can be implemented with any other kind of pulse power source. The voltage across the terminals of the capacitor bank is given by standard circuit theory

$$V(t) = V_0 - C_0^{-1}\int_0^t I(t')dt' - L_0\frac{dI(t)}{dt} - R_0 I(t) \tag{110}$$

which can be reduced using (99) to

$$V(t) = V_0\left(\alpha - \varepsilon\tau - \varepsilon^{-1}\tilde{I}(\tau)\frac{d\tilde{I}(\tau)}{d\tau} - \gamma\tilde{I}(\tau)\right) \tag{111}$$

The voltage across the plasma focus given by (107) is in series with the voltage across the capacitor bank giving a loop voltage equal to zero. This can be rearranged as

$$\tilde{I}(\tau)\frac{d}{d\tau}\{(1+\kappa\mathfrak{L}(\tau))\tilde{I}(\tau)\} = \varepsilon(\alpha - \varepsilon\tau - \gamma\tilde{I}(\tau)) \tag{112}$$

Multiplying both sides by $(1+\kappa\mathfrak{L}(\tau)\tilde{I}(\tau))$ and introducing $\Phi(\tau) \equiv (1+\kappa\mathfrak{L}(\tau))\tilde{I}(\tau)$ this becomes

$$\frac{d\Phi^2(\tau)}{d\tau} = 2\varepsilon\alpha - 2\varepsilon^2\tau + 2(1-\alpha)\varepsilon\kappa\mathfrak{L}(\tau) - 2\varepsilon^2\kappa\tau\mathfrak{L}(\tau) - 2\varepsilon\gamma\Phi(\tau) \tag{113}$$

Equation (113) can be solved by the method of successive approximations by treating $\Phi$ as the limit of a sequence of functions $\{\Phi_0, \Phi_1 \cdots \Phi_n \cdots\}$ for large n, which satisfy the equation

$$\Phi_n^2(\tau) - \Phi_n^2(0) = 2\varepsilon\alpha\tau - \varepsilon^2\tau^2 + 2\alpha\varepsilon\kappa\mathfrak{M}_0(\tau) - 2\varepsilon^2\kappa\mathfrak{M}_1(\tau) - 2\varepsilon\gamma\int_0^\tau \Phi_{n-1}(\tau')d\tau'$$

$$\mathfrak{M}_0(\tau) \equiv \int_0^\tau \mathfrak{L}(\tau)d\tau;\ \mathfrak{M}_1(\tau) \equiv \int_0^\tau \tau\mathfrak{L}(\tau)d\tau \tag{114}$$

where $\Phi_0(\tau)$ is given by

$$\Phi_0^2(\tau) - \Phi_0^2(0) = 2\varepsilon\alpha\tau - \varepsilon^2\tau^2 + 2\alpha\varepsilon\kappa\mathfrak{M}_0(\tau) - 2\varepsilon^2\kappa\mathfrak{M}_1(\tau) \tag{115}$$

The integration constant $\Phi_n^2(0) = \Phi_0^2(0)$ can be calculated using (12):

$$\Phi_0^2(0) = (1 + \kappa\mathfrak{L}(0))^2 \tilde{I}^2(0) = I^2(t_L)/I_0^2 = \frac{4\pi^2 \rho_0 L_0 R_{ins}^2 v_{LB}^2}{\mu_0 \left(\frac{1}{2}C_0 V_0^2\right)} \tag{116}$$

It is shown below that the sequence converges at the 3$^{rd}$ iteration for the illustration considered: the dimensionless current profile $\tilde{I}_n(\tau) = \Phi_n(\tau)/(1 + \kappa\mathfrak{L}(\tau))$ is equal for n=2 and 3.

The inductance function $\mathfrak{L}(\tau)$ can be calculated from the GV surfaces as a function of $\tau$ as shown below, where it would be clear that $\mathfrak{L}(0) = 0$. The solution of (114) then reduces to a quadrature and provides model closure.

For evaluating the inductance, the inverse of the anode profile definition is used, giving the axial coordinate on the anode profile as a function of radial coordinate, $\tilde{Z}_A(\tilde{r})$, instead of the anode radius function $\tilde{R}_A(\tilde{Z})$ given in (1). Then,

$$\mathfrak{L}(\tau) \equiv \int_{\Omega(\tau)} \frac{d\tilde{r}d\tilde{z}}{\tilde{r}} = \int_{\tilde{r}_{min}}^{R_c} d\tilde{r} \frac{\left(\tilde{z}_{GV}(\tilde{r},\tau) - \tilde{Z}_A(\tilde{r})\right)}{\tilde{r}} \tag{117}$$

As an example, for the specific anode profile (83) shown in Fig 6

$$\begin{aligned}
\tilde{Z}_A(\tilde{r}) &= 0 & 1 \leq \tilde{r} \leq \tilde{r}_C \\
&= \tilde{z}_A - \sqrt{\tilde{r}_1^2 - (1-\tilde{r})^2} & 1 - \tilde{r}_1 \leq \tilde{r} < 1 \\
&= \tilde{z}_A + \delta \cdot \frac{(\tilde{r} - (1-\tilde{r}_1))}{(\tilde{r}_2 + \tilde{r}_h - 1 - \tilde{r}_1)} & \tilde{r}_2 + \tilde{r}_h \leq \tilde{r} < (1-\tilde{r}_1) \\
&= \tilde{z}_A + \delta & 0 \leq \tilde{r} < \tilde{r}_2 + \tilde{r}_h
\end{aligned} \tag{118}$$

Fig 16 shows the calculated set of GV surfaces and machine parameters for a transportable plasma focus reported by Rishi Verma et. al. [48], as well as other numbers used in this model (see the profile (83)). Parameters used for algorithmic convenience are: nominal taper δ=0.01, nominal fillet radius $\tilde{r}_1 = \tilde{r}_2 = 0.01$.

| Parameter | Value | Unit | Normalized |
|---|---|---|---|
| Anode radius | 20 | mm | 1 |
| Anode length | 150 | mm | $\tilde{z}_A = 7.5$ |
| Cathode radius | 50 | mm | $\tilde{r}_C = 2.5$ |
| Insulator radius | 20 | mm | $\tilde{r}_I = 1.0$ |
| Insulator length | 30 | mm | $\tilde{z}_I = 1.5$ |
| Cavity radius | 16 | mm | $\tilde{r}_h = 0.8$ |
| Cavity depth | 60 | mm | $\tilde{z}_h = 3$ |
| Capacitance $C_0$ | 50 | μF | |
| Voltage $V_0$ | 17 | kV | |
| Charge $C_0 V_0$ | 0.85 | C | |
| $E_0 = \frac{1}{2} C_0 V_0^2$ | 7.2 | kJ | |
| $L_0$ (measured) | 44 | nH | |
| $R_0$ (measured) | 10 | mΩ | |
| $I_0 = V_0 \sqrt{C_0 / L_0}$ | 573 | kA | |
| $t_{1/4} = \frac{1}{2} \pi \sqrt{L_0 C_0}$ | 2.33 | μs | |
| $\lambda = \frac{1}{2} R_0 \sqrt{C_0 / L_0}$ | 0.1685 | | |
| $I_{max} = I_0 \exp\left(-\lambda \operatorname{ArcCos}(\lambda)/\sqrt{1-\lambda^2}\right)$ | 450 | kA | |
| $p_{max} = 1.05 \times 10^5 f_{LB}^{-2} \left(I_{max}(MA)/R_{ins}(mm)\right)^2$ | $25 f_{LB}^{-2}$ | mbar | |
| Pressure p | 5 | mbar | |
| $\rho_0 = 0.17846 (kg/m^3) p(bar\, D_2)$ | 8.92E-4 | kg/m³ | |
| $Q_m = \mu_0^{-1} \pi a^2 \sqrt{2\mu_0 \rho_0}$ | 0.047 | C | |
| $\varepsilon = Q_m / C_0 V_0$ | 0.056 | | |
| $\kappa = \mu_0 a / 2\pi L_0$ | 0.09 | | |
| $\tau_p = 2(\tilde{z}_A - \tilde{z}_I) + 1$ | 13 | | |
| $I(t_L) = v_{LB} 2\pi R_{ins} \mu_0^{-1} \sqrt{2\mu_0 \rho_0}$ | $137.3 \cdot f_{LB}$ | kA | |
| $\eta_C = (1 - \varepsilon \tau_p)^2$ | 0.08 | | |

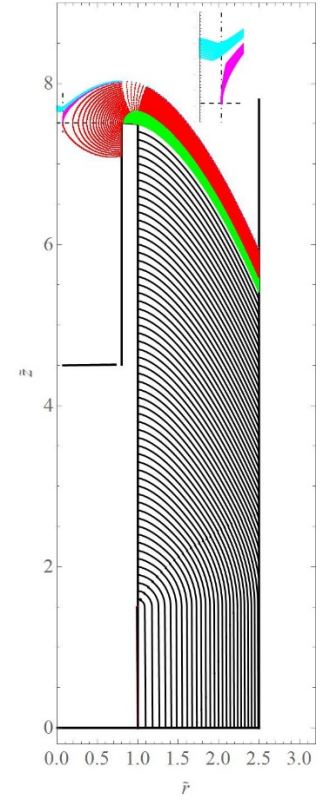

Fig 16: GV surfaces and machine parameters for a transportable 10 kJ plasma focus [48]. The rundown phase is shown in black, turn-around at the first fillet and up to the beginning of second fillet is shown in green, the propagation up to $\tilde{r}_{int}$ is shown in red and the formation of zero velocity boundary is shown in magenta and cyan.

The inductance function $\mathcal{L}(\tau)$ is shown in Fig 17. The red and blue coloured portions show the inductance with and without the contribution of the flux in the cavity.

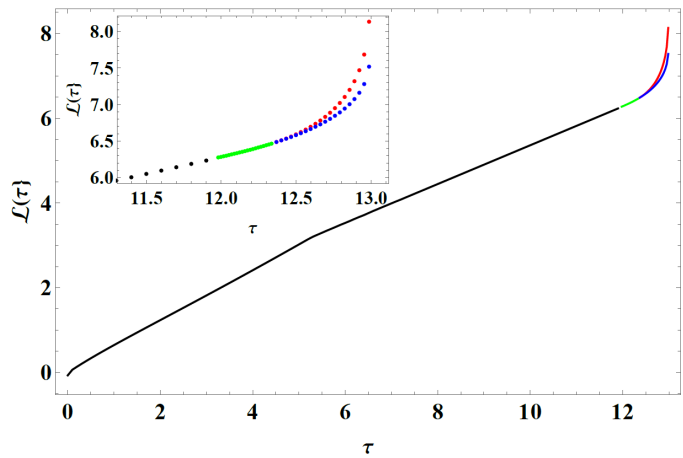

Fig 17: Dimensionless inductance function $\mathcal{L}(\tau)$. The black, green and red portions correspond to the inductance contribution from the similarly coloured GV surfaces in Fig 16. The inset shows an expanded view of the function. The blue dots show the inductance variation if the flux in the cavity is ignored, i.e. with a solid anode.

Fig 18 shows how the iterations converge.

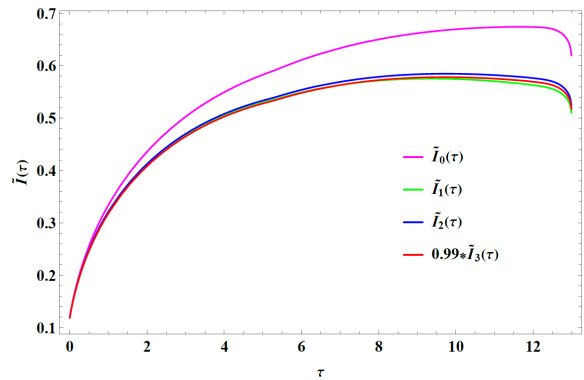

Fig 18. Convergence of iterations for the current profile for the plasma focus shown in Fig 16.

Fig 19 compares the calculated waveform with the experimental waveform, both scaled to their respective maximum values.

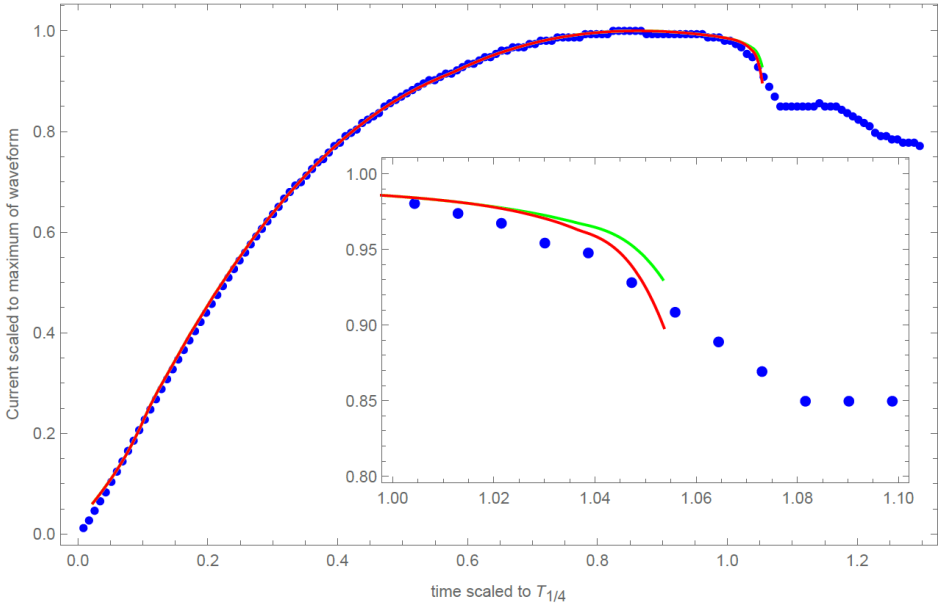

Fig. 19: Comparison of experimental waveform with calculated waveforms. The experimental waveform (blue dots) is calculated by numerical integration from uncalibrated current derivative waveform data. Its ordinate is scaled to maximum of its amplitude and abscissa with the quarter cycle time. The red and green curves are similarly scaled calculated waveforms, with and without the flux in the cavity. The constants $L_0$ and $R_0$ are found by fitting another uncalibrated current waveform at 30 mbar of argon to equation (14). The pressure value is measured using a capacitance manometer. The only fitting parameter used is $f_{LB} = 0.15$, which gives $t_L = 54$ ns and $I(t_L) = 20.6 \text{kA}$. Data kindly provided by Dr. Rishi Verma.

Unlike the RGV model, this current calculation uses experimentally determined values of $L_0$, $R_0$ and pressure. The only parameter value obtained by fitting is $f_{LB} = 0.15$, which is expected to be a material constant for deuterium, determined once for all. Putting $f_{LB} = 0.1$ and $f_{LB} = 0.2$ translates the calculated waveforms to higher and lower sides of the time axis respectively. With the lower value of the threshold parameter $f_{LB}$, an artificial current loss factor could restore that shift. The good agreement between calculated and experimental waveforms in the rundown phase bears a striking contrast with the observed deviation in the radial phase. The calculated current waveform in the radial phase has a faster decrease with time as compared with the experimental waveform. This is consistent with the neglect of the poloidal flux emission phenomenon, which diverts a part of input power away from the azimuthal magnetic field. This could also in principle be "corrected" by introducing an artificial current loss factor [50]. Characteristic features of this deviation provide important clues to the construction of a physical theory that evaluates corrections to the power equation (103). This serves as an illustration of the role of the kinematic model as a reference for quantifying the discrepancy between a simple model and experimental data that should point the way towards further refinement of the model into a closer approximation to a physical theory.

The closure of the kinematic model is represented by recovery of real time t from the dimensionless time τ according to (98). This is shown in Fig 20.

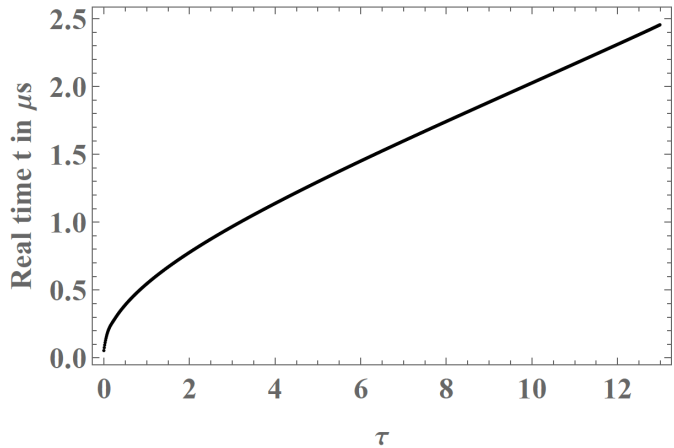

Fig 20: Recovery of real time in μs from the dimensionless time, representing model closure,

With this step, the variation of scaling parameters introduced in Section IIA in space and time stands fully determined, providing closure to the kinematic model. The illustrated procedure is fairly general and can be carried out with any shape of the anode and insulator, any working gas and any well-characterized pulse power source.

This exercise also brings out two important points:

(a) The kinematic model can predict the experimental waveform in the rundown phase with measured values of the static inductance, static resistance and fill pressure of deuterium without the use of fitted parameters, once the value of $v_{LB}$ is confirmed as a material constant. This is sufficient to serve as a reliable design tool for conceptualizing new plasma focus facilities for maximum energy transfer from the energy storage to the pinch, where the use of parameters fitted to a current waveform is not possible.

(b) A theory of the plasma focus should aim to resolve the discrepancy between the calculated and measured current waveforms in the radial phase by determining corrections to the inductance model from physical phenomena ignored in the kinematic model.

## VI. Summary and Conclusion

This paper is Part I of a planned series of papers aimed at development of a theory of the Dense Plasma Focus. This program begins by introducing the proposition that a physical theory of the plasma focus needs to be built in an incremental manner, as a hierarchy of partial theories which progressively relax a corresponding hierarchy of simplifying assumptions. The most basic level in such a hierarchy is proposed to be a kinematic model that represents the shape and location of the boundaries of the propagating plasma focus sheath by a family of 3-D surfaces of rotation. The physical basis of this kinematic model rests on the scaling properties of the equations of motion of the plasma in its single-fluid continuum approximation. Time and space dependent scaling parameters characteristic of the plasma focus reduce these equations of motion into a machine-independent dimensionless form. The task of developing a theory of the plasma focus can then be reduced to two sub-problems: (1) determination of the scaling parameters as a function of space and time leading to a family of computed surfaces of rotation whose shapes are known to resemble the observed shape of the plasma [17] (2) Determination of the variation of the scaled parameters in space and time in the vicinity of these surfaces of rotation. The first task forms the subject of this Part I.

This kinematic model is a generalization of the Gratton-Vargas model [1] and its revised form, the Resistive Gratton-Vargas model [5,6], that includes the following additional features:

1) It takes into account the time interval between the start of current and start of plasma propagation in terms of a threshold value of the scaling velocity related to the specific energy for dissociation and ionization of the fill gas.

2) It introduces a generalized procedure for computing the 3-D surfaces of rotation resembling the shape of the plasma for mathematically defined profiles of anode and insulator.

3) It represents the front and rear boundaries of the plasma sheath in terms of the computed surfaces of rotation. It borrows the basic idea of Potter's slug model [40] that the zero-velocity boundary of a z-pinch is formed by a collision of the front boundary of the plasma reflected from the axis with the imploding rear boundary. This allows it to compute the evolution of the shape of the plasma focus in its pinch phase. It is able to reproduce some experimentally observed numbers related to the pinch phase.

4) It shows formation of some bounded structures embedded within the plasma stem, mimicking similar observations reported from experiments, which are believed [4] to play a central role in production of neutrons.

5) It introduces a generalized approach to computing the current waveform for a plasma focus powered by any kind of pulsed power source. The usual circuit representation of a plasma focus as a time-varying inductance is shown to be a special case of a more general formalism based on first principles. This special case neglects dynamo effects that convert plasma kinetic energy into magnetic energy using the geomagnetic field as a seed. Existence of this effect has recently been demonstrated by detection of poloidal magnetic flux emission [49] from a plasma focus before, during and after the pinch phase.

Comparison of computed current waveform with experimentally determined current waveform shows good agreement with the rundown phase with only one adjustable parameter that should be a material constant for deuterium, to be determined once and for all. This implies that the energy transfer efficiency from the power source to the pinch phase can be predicted without introducing any fitting parameters. This enables the present theory to perform as a reliable design tool for new plasma focus facilities, which is one of the major objectives of this exercise.

Deviation of the computed current waveform from the experimental one in the radial phase is consistent with the first principles theory [19] of the representation of the plasma focus as a circuit element. This deviation provides important clues to the development of a physical theory that accounts for phenomena ignored in the kinematic model.

With this, the first stage of laying the kinematic foundation of a theory of the Dense Plasma Focus is complete. A physical theory that accounts for the temporal evolution of the spatial distribution of scaled parameters velocity $\tilde{v}$, density $\tilde{\rho}$, magnetic field $\tilde{B}$ and pressure $\tilde{p}$ in the vicinity of the GV surfaces is the next objective.

Acknowledgements: Kind help of Dr. Rishi Verma of Bhabha Atomic Research Centre is gratefully acknowledged.